\documentclass[12pt]{article}

\usepackage{epsfig}

\def\comp{{\rm C}\llap{\vrule height7.1pt width1pt depth-.4pt\phantom t}}
\def\Fint{\rlap{$\Biggl\rfloor$}\Biggl\lceil}
\def\square{\kern1pt\vbox{\hrule height 1.2pt\hbox{\vrule width 1.2pt\hskip 3pt
   \vbox{\vskip 6pt}\hskip 3pt\vrule width 0.6pt}\hrule height 0.6pt}\kern1pt}
\def\gtwid{\mathrel{\raise.3ex\hbox{$>$\kern-.75em\lower1ex\hbox{$\sim$}}}}
\def\ltwid{\mathrel{\raise.3ex\hbox{$<$\kern-.75em\lower1ex\hbox{$\sim$}}}}
\def\comp{{\rm C}\llap{\vrule height7.1pt width1pt depth-.4pt\phantom t}}
\def\ltwid{\mathrel{\raise.3ex\hbox{$<$\kern-.75em\lower1ex\hbox{$\sim$}}}}
\def \be{\begin{equation}}
\def \ee{\end{equation}}
\def \bea{\begin{eqnarray}}
\def \eea{\end{eqnarray}}

\def \del{\partial}

\def \f{\frac}

\begin{document}

\begin{titlepage}

\begin{flushright}
SPIN-08/11, ITP-UU-08/12 \\ UFIFT-QG-08-02
\end{flushright}

\vskip 1cm

\begin{center}
{\bf A Simple Operator Check of the Effective Fermion Mode Function
during Inflation}
\end{center}

\vskip .3cm

\begin{center}
S. P. Miao$^*$
\end{center}

\begin{center}
\it{Institute for Theoretical Physics \& Spinoza Institute, Utrecht University\\
Leuvenlaan 4, Postbus 80.195, 3508 TD Utrecht, THE NETHERLANDS}
\end{center}

\vskip .3cm

\begin{center}
and
\end{center}

\vskip .3cm

\begin{center}
R. P. Woodard$^{\dagger}$
\end{center}

\begin{center}
\it{Department of Physics, University of Florida \\
Gainesville, FL 32611, UNITED STATES}
\end{center}

\vspace{.5cm}

\begin{center}
ABSTRACT
\end{center}

We present a relatively simple operator formalism which reproduces
the leading infrared logarithm of the one loop quantum gravitational
correction to the fermion mode function on a locally de Sitter
background. This rule may serve as the basis for an eventual
stochastic formulation of quantum gravity during inflation. Such a
formalism would not only effect a vast simplification in obtaining
the leading powers of $\ln(a)$ at fixed loop orders, it would also
permit us to sum the series of leading logarithms. A potentially
important point is that our rule does not seem to be consistent with
any simple infrared truncation of the fields. Our analysis also
highlights the importance of spin as a gravitational interaction
that persists even when kinetic energy has redshifted to zero.

\begin{flushleft}
\end{flushleft}

\begin{flushleft}
$^*$ e-mail: S.P.Miao@phys.uu.nl \\
$^{\dagger}$ e-mail: woodard@phys.ufl.edu
\end{flushleft}

\end{titlepage}

\section{Introduction}
Gravitons and massless, minimally coupled (MMC) scalars are unique
in being massless without classical conformal invariance. The
combination of these properties causes the accelerated expansion of
spacetime during inflation to tear long wavelength virtual quanta
out of the vacuum \cite{TW3,TW5RPW1}. As more and more gravitons and
MMC scalars emerge from the vacuum, the metric and MMC scalar field
strengths experience a slow growth. The effect can be felt by any
quantum field theory which involves either the undifferentiated
metric or an undifferentiated MMC scalar.

An example is the one loop enhancement recently found \cite{MW2,MW3}
for the plane wave mode functions of massless, Dirac fermions which
are coupled to quantum gravity on a locally de Sitter background,
\begin{equation}
ds^2 = -dt^2 + a^2(t) d\vec{x} \!\cdot\! d\vec{x} \qquad {\rm where}
\qquad a(t) = e^{H t} \; .\label{deSitter}
\end{equation}
(This background solves the classical Friedmann equation, $(D-1)
(\dot{a}/a)^2 = \Lambda \equiv (D-1) H^2$, where $\Lambda$ is the
cosmological constant. The one-loop back-reaction on $a(t)$ cannot
affect the fermion mode function until two loop order.) At late
times the full mode function $\Xi(x;\vec{k},s)$ behaves as if the
tree order mode function $\Xi_0(x;\vec{k},s)$ was subject to a
time-dependent field strength renormalization,
\begin{equation}
\Xi(x;\vec{k},s) \longrightarrow
\frac{\Xi_0(x;\vec{k},s)}{\sqrt{Z_2(t)}} \; .
\end{equation}
This field strength renormalization takes the form,
\begin{equation}
Z_2(t) = 1 - \frac{17}{4\pi} G H^2 \ln(a) + O(G^2) \; , \label{Z(t)}
\end{equation}
where $G$ and $H$ are the Newton and Hubble constants, respectively.

The factor of $\ln(a) = H t$ in expression (\ref{Z(t)}) is known as
an {\it infrared logarithm}. Any quantum field theory which involves
undifferentiated MMC scalars or metrics will show similar infrared
logarithms in some of its Green's functions. They arise at one and
two loop orders in the expectation value of the stress tensor and in
the scalar self-mass-squared of a MMC scalar with a quartic
self-coupling \cite{OWBK}. In scalar quantum electrodynamics they
have been seen in the one loop vacuum polarization \cite{PTWP} and
the two loop expectation values of scalar bilinears \cite{PTsW1},
the field strength bilinear and the stress tensor \cite{PTsW2}. In
Yukawa theory they show up in the one loop fermion self-energy
\cite{PW2GP} and in the two loop coincident vertex function
\cite{MW1}. In pure quantum gravity they occur in the one loop
graviton self-energy \cite{TW0} and in the two loop expectation
value of the metric \cite{TW1}. They even contaminate loop
corrections to the power spectrum of cosmological perturbations
\cite{SW1,BSVMSKCBPvdmsUM} and other fixed-momentum correlators
\cite{SW2}.

Infrared logarithms introduce a fascinating secular element into the
usual, static results of quantum field theory. Their most intriguing
property is their ability to compensate for powers of the loop
counting parameter which suppress quantum loop effects. Indeed, the
continued growth of $\ln(a) = H t$ must eventually {\it overwhelm}
the loop counting parameter, no matter how small it is. However,
this does not necessarily mean that quantum loop effects become
strong. The correct conclusion is rather that perturbation theory
breaks down past a given point in time. One must employ a
nonperturbative technique to follow what happens later.

Certain models lend themselves to resummation schemes such as the
$1/N$ expansion \cite{CMBCVHSSRS}, but a more general technique is
suggested by the form of the expansion for $Z_2(t)$ in (\ref{Z(t)}),
\begin{equation}
Z_2(t) = 1 + \sum_{\ell=1}^{\infty} (G H^2)^{\ell} \Biggl\{
c_{\ell,0} \Bigl[\ln(a)\Big]^{\ell} + c_{\ell,1}
\Bigl[\ln(a)\Bigr]^{\ell-1} + \dots + c_{\ell,\ell-1} \ln(a)\Biggr\}
. \label{genform}
\end{equation}
Here the constants $c_{\ell,k}$ are pure numbers which are assumed
to be of order one. The term in (\ref{genform}) involving $[G H^2
\ln(a)]^{\ell}$ is the {\it leading logarithm} contribution at
$\ell$ loop order; the other terms are {\it subdominant logarithms}.
Perturbation theory breaks down when $\ln(a) \sim 1/{G H^2}$, at
which point the leading infrared logarithms at each loop order
contribute numbers of order one. In contrast, the subleading
logarithms are all suppressed by at least one factor of the small
parameter $G H^2 \ltwid 10^{-12}$. So it makes sense to retain only
the leading infrared logarithms,
\begin{equation}
Z_2(t) \longrightarrow 1 + \sum_{\ell=1}^{\infty} c_{\ell,0} \Bigl[G
H^2 \ln(a)\Bigr]^{\ell} \; .
\end{equation}
This is known as the {\it leading logarithm approximation}.

Starobinski\u{\i} has developed a simple stochastic formalism
\cite{AAS} which reproduces the leading infrared logarithms at each
order \cite{RPW3} for any scalar potential model of the form,
\begin{equation}
\mathcal{L} = -\frac12 \partial_{\mu} \varphi \partial_{\nu} \varphi
g^{\mu\nu} \sqrt{-g} - V(\varphi) \sqrt{-g} \; . \label{Starform}
\end{equation}
Probabilistic representations of inflationary cosmology have been
much studied in order to understand initial conditions \cite{AVNS}
and global structure \cite{GLMLM}. More recently they have been
employed to study non-Gaussianity \cite{RST}. However, we wish here
to focus on Starobinski\u{\i}'s technique as a wonderfully simple
way of recovering the most important secular effects of inflationary
quantum field theory \cite{SJRSNNWVMMENPR}. It is of particular
importance for us that Starobinski\u{\i} and Yokoyama have shown how
to take the late time limit of the series of leading infrared
logarithms whenever the potential $V(\varphi)$ is bounded below
\cite{SY}. This is the true analogue of what the renormalization
group accomplishes in flat space quantum field theory and
statistical mechanics.

The solution of Starobinski\u{\i} and Yokoyama is an amazing
achievement, but it only gives us control over infrared logarithms
which arise in scalar potential models (\ref{Starform}). The most
general theories which show infrared logarithms possess two
complicating features:
\begin{itemize}
\item{Couplings to fields other than MMC scalars and gravitons; and}
\item{Interactions which involve differentiated MMC scalars and
gravitons.\footnote{Of course there would be no infrared logarithms
if {\it all} the MMC scalars and gravitons were differentiated.
However, infrared logarithms must arise, in the expectation values
of some operators, from interactions which involve at least one
undifferentiated MMC scalar or graviton. Examples include the $h^n
\partial h \partial h$ interaction of pure quantum gravity
\cite{TW0,TW1} and scalar interactions of the form $\varphi^2
\partial \varphi \partial \varphi$ \cite{SW1,RPW3}.}}
\end{itemize}
An important step forward was the recent leading log solutions for
MMC scalars which are either Yukawa-coupled to a massless, Dirac
fermion \cite{MW1} or to electrodynamics \cite{PTsW3}. Although the
second model has derivative interactions, this feature was avoided
(at leading logarithm order) by working in Lorentz gauge. We still
do not understand how to treat derivative interactions.

At the level of dimensionally regularized perturbation theory, the
scalar leading logarithm solutions which have so far been obtained
can be reduced to five simple steps \cite{PTsW3}:
\begin{enumerate}
\item{Expand the full scalar operator $\varphi(x)$ in powers of the free
field $\varphi_0(x)$ which agrees with $\varphi(x)$ and its first
derivative at the beginning of inflation as described in the recent
paper by Musso \cite{Musso};}
\item{The expectation value of any desired operator can then be expressed
as vertex integrations of retarded Green's functions times products
of expectation values of pairs of free fields;}
\item{Make the following replacement for the expectation value of two free
fields:
\begin{equation}
\Bigl\langle \Omega \Bigl\vert \varphi_0(x) \varphi_0(x') \Bigr\vert
\Omega \Bigr\rangle \longrightarrow \frac{H^{D-2}}{(4
\pi)^{\frac{D}2}} \frac{\Gamma(D\!-\!1)}{\Gamma(\frac{D}2)} \, 2
\ln\Bigl[{\rm min}(a,a') \Bigr] \; ; \label{srule1}
\end{equation}}
\item{Make the following replacement for the retarded Green's function:
\begin{equation}
G(x;x') \longrightarrow \frac{\theta(t\!-\!t') \delta^{D-1}(\vec{x}
\!-\! \vec{x}')}{(D\!-\!1) H a^{\prime D-1}} \; ; {\rm and}
\label{srule2}
\end{equation}}
\item{Evaluate the contributions from any other fields (for example,
photons or fermions) exactly to the required order.}
\end{enumerate}
Indeed, these rules even predict the occasional null results
\cite{DWKW1} that sometimes occur at low orders.

It is straightforward to show that this old rule
(\ref{srule1}-\ref{srule2}) does {\it not} suffice to recover
(\ref{Z(t)}). The purpose of this paper is to devise a simple rule
which does work. We do not yet know if this rule applies either to
other quantities or to higher loop orders. Nor do we possess a
nonperturbative realization for this rule. Our rule nonetheless
represents very significant progress in the struggle to solve
inflationary quantum gravity in the leading logarithm approximation.
Such a solution would make it simple to compute leading logarithm
results at fixed order, and would also facilitate summation of the
series of leading logarithms, thereby defining evolution past the
breakdown of perturbation theory.

In section 2 we explain how solving the Schwinger-Keldysh effective
field equations is equivalent to computing the expectation value of
a suitable canonical operator. Section 3 works out the operator and
its expectation value to the order we require. At this stage the
result is still exact and represents no simplification of the
effective field equation technique. Our simplifying rule is
presented in section 4. In section 5 we demonstrate that the rule
indeed reproduces the leading infrared logarithm in the one loop
correction to the fermion mode function. Our conclusions comprise
section 6.

\section{The Effective Mode Function}
We begin this section by describing the Schwinger-Keldysh formalism.
This is a covariant extension of Feynman diagrams that produces true
expectation values instead of in-out matrix
elements\cite{JS,KTM,BM,LVK}. We then review the quantum-corrected
Dirac equation whose solution (for spatial plane waves) gives the
$\comp$-number effective fermion mode function $\Xi_i(x;\vec{k},s)$.
The section closes by giving the connection between
$\Xi_i(x;\vec{k},s)$ and the fer\-mi\-on operator $\Psi_i(x)$.

The in-out effective field equations give a fine representation of
flat space scattering problems but they are not typically suitable
for cosmological settings in which particle production precludes the
in vacuum from evolving to the out vacuum. Persisting with thein-out
formalism on de Sitter background would result in processes being
dominated by infrared divergences from the enormous spacetime volume
of the infinite future \cite{TW3,TW4}. The better course in this
case is to release the universe in a prepared state at finite time
and let it evolve as it will. Problems of this sort are described by
the Schwinger-Keldysh effective field equations \cite{RJCSHYCH}.

Consider a scalar field $\varphi$ whose Lagrangian (by which we mean
the spatial integral of the Lagrangian density) at time $t$ is
$L[\varphi(t)]$. The fundamental relation between the canonical
operator formalism and the Schwinger-Keldysh functional integral
formalism is \cite{FW},
\begin{eqnarray}
\lefteqn{\Bigl\langle \Phi \Bigl\vert
\overline{T}\Bigl(\mathcal{O}_2[ \varphi]\Bigr)
T\Bigl(\mathcal{O}_1[\varphi]\Bigr) \Bigr\vert \Phi \Bigr\rangle =
\Fint [d\varphi_{\scriptscriptstyle \!+}]
[d\varphi_{\scriptscriptstyle \!-}] \,
\delta\Bigl[\varphi_{\scriptscriptstyle \!-}\!(t_1) \!-\! \varphi_{
\scriptscriptstyle \!+}\!(t_1)\Bigr]
\Phi^*[\varphi_{\scriptscriptstyle -}\!
(t_0)] \Phi[\varphi_{\scriptscriptstyle \!+}\!(t_0)] } \nonumber \\
& & \hspace{3.5cm} \times \mathcal{O}_2[\varphi_{\scriptscriptstyle
\!-}] \mathcal{O}_1[\varphi_{\scriptscriptstyle \!+}] \exp\Biggl[i
\!\int_{t_0}^{t_1} \!\!\!dt \, \Bigl\{L[\varphi_{\scriptscriptstyle
\!+}\!(t)] - L[\varphi_{ \scriptscriptstyle \!-} \!(t)]\Bigr\}
\Biggr] \; . \quad \label{fund}
\end{eqnarray}
Here $\vert \Phi \rangle$ is the Heisenberg state whose wave
functional in terms of the $\varphi$ eigenkets at time $t_0$ is
$\Phi[\varphi(t_0)]$. The canonical expectation value on the left
hand side consists of the product of an anti-time-ordered operator
$\mathcal{O}_2[\varphi]$ times a time-ordered operator
$\mathcal{O}_1[\varphi]$. The value of $t_1
> t_0$ is arbitrary as long as it is in the future of the latest operator
occurring in either $\mathcal{O}_1$ or $\mathcal{O}_2$.

The Feynman rules follow from relation (\ref{fund}) in close analogy
to those for in-out matrix elements. Because the same field is
represented by two different dummy functional variables,
$\varphi_{\scriptscriptstyle \!\pm}\!(x)$, the endpoints of lines
carry a $\pm$ polarity. External lines associated with the
anti-time-ordered operator $\mathcal{O}_2[\varphi]$ have the $-$
polarity whereas those associated with the time-ordered operator
$\mathcal{O}_1[\varphi]$ have the $+$ polarity. Interaction vertices
are either all $+$ or all $-$. Vertices with $+$ polarity are the
same as in the usual Feynman rules whereas vertices with the $-$
polarity have an additional minus sign. Propagators can be $++$,
$-+$, $+-$ or $--$.

From this sketch we see that the N-point one-particle-irreducible
(1PI) function $\Gamma^N(x_1,\dots,x_N)$ of the in-out formalism
gives rise to $2^N$ different Schwinger-Keldysh 1PI functions
$\Gamma^N(x_{1\pm}, \dots,x_{N\pm})$. Now recall that the in-out
effective action is the generating functional of in-out 1PI
functions,
\begin{equation}
\Gamma[\phi] = \sum_N \frac1{N!} \int d^4x_1 \phi(x_1) \dots \int
d^4x_N \phi(x_N) \times \Gamma^N(x_1,\dots,x_N) \; .
\end{equation}
The analogous generating functional for Schwinger-Keldysh 1PI
functions is,
\begin{equation}
\Gamma[\phi_{\scriptscriptstyle \!+},\phi_{\scriptscriptstyle \!-}]
= \sum_N \frac1{N!} \int d^4x_1 \phi_{\scriptscriptstyle
\!\pm}\!(x_1) \dots \int d^4x_N \phi_{\scriptscriptstyle \!\pm}(x_N)
\times \Gamma^N(x_{1\pm},\dots,x_{N\pm}) \; .
\end{equation}
The Schwinger-Keldysh effective field equations are obtained by
varying this functional with respect to either
$\phi_{\scriptscriptstyle \!+}$ or $\phi_{\scriptscriptstyle \!-}$,
and then setting the two fields equal,
\begin{equation}
\frac{\delta \Gamma[\phi_{\scriptscriptstyle
\!+},\phi_{\scriptscriptstyle \!-}]}{\delta \phi_{\scriptscriptstyle
\!+}\!(x)} \Biggl\vert_{ \phi_{\scriptscriptstyle \!\pm} = \phi} = 0
\; .
\end{equation}

It is worth being a little more explicit for the case in which the
$0$-point and $1$-point functions vanish. If the classical action is
$S[\phi]$, and the self-mass-squared is $-i M^2_{\scriptscriptstyle
\!\pm\pm}\!(x;x')$, the Schwinger-Keldysh effective action has the
following expansion,
\begin{eqnarray}
\lefteqn{\Gamma[\phi_{\scriptscriptstyle
\!+},\phi_{\scriptscriptstyle \!-}] = S[\phi_{\scriptscriptstyle
\!+}] - S[\phi_{\scriptscriptstyle \!-}]
-\frac12 \int \!d^4x \int d^4x' } \nonumber \\
& & \hspace{1cm} \times \left\{\matrix{ \phi_{\scriptscriptstyle
\!+}\!(x) M^2_{\scriptscriptstyle \!++}\!(x;x')
\phi_{\scriptscriptstyle \!+}\!(x') + \phi_{\scriptscriptstyle
\!+}\!(x) M^2_{\scriptscriptstyle \!+-}\!(x;x')
\phi_{\scriptscriptstyle \!-}\!(x') \cr + \phi_{\scriptscriptstyle
\!-}\!(x) M^2_{\scriptscriptstyle \!-+}\!(x;x')
\phi_{\scriptscriptstyle \!+}\!(x') + \phi_{\scriptscriptstyle
\!-}\!(x) M^2_{\scriptscriptstyle \!--}\!(x;x')
\phi_{\scriptscriptstyle \!-}\!(x') \cr}\right\} + O(\phi^3) \; .
\qquad
\end{eqnarray}
The Schwinger-Keldysh effective field equations are,
\begin{equation}
\frac{\delta S[\phi]}{\delta \phi(x)} - \int d^4x' \Bigl[
M^2_{\scriptscriptstyle \!++}\!(x;x') + M^2_{\scriptscriptstyle
\!+-}\!(x;x')\Bigr] \phi(x') + O(\phi^2) = 0 \; . \label{SKEE}
\end{equation}
The quantum-corrected Klein-Gordon equation results from linearizing
(\ref{SKEE}), and its solution for a spatial plane wave is the
scalar effective mode function. The peculiar combination of $
M^2_{\scriptscriptstyle \!++}\!(x;x') + M^2_{\scriptscriptstyle
\!+-}\!(x;x')$ in (\ref{SKEE}) has two important properties:
\begin{itemize}
\item{It is real, even though each self-mass-squared has a nonzero
imaginary part; and}
\item{It vanishes for any point $x^{\prime \mu}$ outside the past
light-cone of $x^{\mu}$.}
\end{itemize}

This paper concerns our solution of the quantum-corrected Dirac
equation for the effective fermion mode function \cite{MW3},
\begin{equation}
i \hspace{-.1cm}\not{\hspace{-.10cm} \mathcal{\partial}}_{ij}
\Xi_j(x) = \int d^4x' \, \Biggl\{ \Bigl[\mbox{}_i \Sigma_j
\Bigr]_{\scriptscriptstyle \!++}\!\!(x;x') + \Bigl[\mbox{}_i
\Sigma_j \Bigr]_{\scriptscriptstyle \!+-}\!\!(x;x') \Biggr\} \,
\Xi_j(x) \; . \label{Diraceqn}
\end{equation}
Here $\hspace{-.1cm} \not{\hspace{-.1cm} \partial}_{ij} \equiv
\gamma^{\mu}_{ij} \partial_{\mu}$ and $\gamma^{\mu}_{ij}$ represents
the usual, $4 \times 4$ gamma matrices. Note that $\Xi_i(x)$ is a
4-component $\comp$-number field, even though the associated
canonical operator $\Psi_i(x)$ is fermionic. There is no trace of
the de Sitter geometry in the classical part of (\ref{Diraceqn})
because we work in conformal coordinates,
\begin{equation}
ds^2 = a^2 \Bigl[ -d\eta^2 + d\vec{x} \cdot d\vec{x}\Bigr] \equiv
a^2 \eta_{\mu\nu} dx^{\mu} dx^{\nu} \quad {\rm where} \quad a =
-\frac1{H \eta} = e^{H t} \; .
\end{equation}
Massless fermions are conformally invariant in any dimension so we
computed the fermion self-energy using dimensional regularization
for the conformally rescaled field,
\begin{equation}
\Psi_i(x) \equiv a^{\frac{D-1}2} \, \psi_i(x) \; .
\end{equation}
This removes any dependence upon the de Sitter scale factor from the
tree order equation for $\Psi_i(x)$, and hence for $\Xi_i(x)$.

Gravity is {\it not} conformally invariant, so one loop quantum
gravitational corrections to the fermion self-energy involve the de
Sitter scale factor. In computing these corrections we fixed the
local Lorentz gauge so as to allow an algebraic expression for the
vierbein in terms of the metric \cite{RPW2}. The general coordinate
gauge was fixed to make the tensor structure of the graviton
propagator decouple from its spacetime dependence \cite{TW2RPW0}.
After absorbing the divergences with three BPHZ
(Bogoliubov-Parasiuk-Hepp-Zimmermann) counterterms we took the
unregulated limit of $D=4$ to obtain the following results
\cite{MW2}:
\begin{eqnarray}
\lefteqn{\Bigl[\Sigma\Bigr]_{\scriptscriptstyle \!++}\!\!(x;x') =
\frac{i \kappa^2 H^2}{2^6 \pi^2} \Biggl\{\frac{\ln(a a')}{H^2 a a'}
\hspace{-.1cm} \not{\hspace{-.1cm} \partial} \partial^2 \!+\!
\frac{15}2 \ln(a a') \hspace{-.1cm} \not{\hspace{-.1cm} \partial}
\!-\! 7 \ln(a a') \; \hspace{-.1cm} \overline{\not{\hspace{-.1cm}
\partial}} \Biggr\}
\delta^4(x \!-\! x') } \nonumber \\
& & \hspace{.2cm} + \frac{\kappa^2}{2^8 \pi^4 a a'} \hspace{-.1cm}
\not{\hspace{-.1cm} \partial} \partial^4 \Bigl[ \frac{\ln(\mu^2
\Delta x^2_{\scriptscriptstyle \!++})}{\Delta x^2_{
\scriptscriptstyle \!++}} \Bigr] + \frac{\kappa^2 H^2}{2^8 \pi^4}
\Biggl\{\Bigl(\frac{15}2 \hspace{-.1cm} \not{\hspace{-.1cm}
\partial} \,
\partial^2 - \hspace{-.1cm} \overline{\not{\hspace{-.1cm} \partial}} \,
\partial^2 \Bigr) \Bigl[ \frac{\ln(\mu^2 \Delta x^2_{\scriptscriptstyle \!++})
}{\Delta x^2_{\scriptscriptstyle \!++}} \Bigr] \nonumber \\
& & \hspace{1.1cm} + \Bigl(-8 \; \hspace{-.1cm}
\overline{\not{\hspace{-.1cm}
\partial}} \partial^2 \!+\! 4 \hspace{-.1cm} \not{\hspace{-.1cm} \partial}
\nabla^2 \Bigr) \Bigl[ \frac{\ln(\frac14 H^2\Delta
x^2_{\scriptscriptstyle \! ++})}{\Delta x^2_{\scriptscriptstyle
\!++}} \Bigr] \!+\! 7 \hspace{-.1cm} \not{\hspace{-.1cm} \partial}
\, \nabla^2 \Bigl[ \frac1{\Delta x^2_{
\scriptscriptstyle \!++}} \Bigr]\!\Biggr\} + O(\kappa^4) \; , \qquad \\
\lefteqn{\Bigl[\Sigma\Bigr]_{\scriptscriptstyle \!+-}\!\!(x;x') =
\frac{-\kappa^2}{2^8 \pi^4 a a'} \hspace{-.1cm} \not{\hspace{-.1cm}
\partial} \partial^4 \Bigl[
\frac{\ln(\mu^2 \Delta x^2_{\scriptscriptstyle \!+-})}{\Delta x^2_{
\scriptscriptstyle \!+-}} \Bigr] \!-\! \frac{\kappa^2 H^2}{2^8
\pi^4} \Biggl\{\!\!\Bigl(\frac{15}2 \hspace{-.1cm}
\not{\hspace{-.1cm}
\partial} \,
\partial^2 \!\!-\! \hspace{-.1cm} \overline{\not{\hspace{-.1cm} \partial}} \,
\partial^2 \Bigr)\!\Bigl[\frac{\ln(\mu^2 \!\Delta x^2_{\scriptscriptstyle \!+-})
}{\Delta x^2_{\scriptscriptstyle \!+-}} \Bigr] } \nonumber \\
& & \hspace{1.1cm} + \Bigl(-8 \; \hspace{-.1cm}
\overline{\not{\hspace{-.1cm}
\partial}} \partial^2 \!+\! 4 \hspace{-.1cm} \not{\hspace{-.1cm} \partial}
\nabla^2 \Bigr) \Bigl[ \frac{\ln(\frac14 H^2\Delta
x^2_{\scriptscriptstyle \! +-})}{\Delta x^2_{\scriptscriptstyle
\!+-}} \Bigr] \!+\! 7 \hspace{-.1cm} \not{\hspace{-.1cm} \partial}
\, \nabla^2 \Bigl[ \frac1{\Delta x^2_{ \scriptscriptstyle \!+-}}
\Bigr]\!\Biggr\} + O(\kappa^4) \; . \qquad
\end{eqnarray}
Here $\kappa^2 \equiv 16 \pi G$ is the loop counting parameter of
quantum gravity. The various differential and spinor-differential
operators are,
\begin{equation}
\partial^2 \equiv \eta^{\mu\nu} \partial_{\mu} \partial_{\nu} \;\; , \;\;
\nabla^2 \equiv \partial_i \partial_i \;\; , \;\; \hspace{-.1cm}
\not{\hspace{-.1cm} \partial} \equiv \gamma^{\mu} \partial_{\mu}
\;\; {\rm and} \;\; \hspace{-.1cm} \overline{\not{\hspace{-.1cm}
\partial}} \, \equiv \gamma^i
\partial_i \; .
\end{equation}
The two conformal coordinate intervals are,
\begin{eqnarray}
\Delta x^2_{\scriptscriptstyle ++}\!(x;x') & \equiv & \Vert \vec{x}
\!-\! \vec{x}'\Vert^2 - \Bigl(\vert\eta \!-\! \eta'\vert - i
\delta\Bigr)^2 \; ,
\label{x++} \\
\Delta x^2_{\scriptscriptstyle +-}\!(x;x') & \equiv & \Vert \vec{x}
\!-\! \vec{x}'\Vert^2 - \Bigl(\eta \!-\! \eta'+ i \delta\Bigr)^2 \;
. \label{x+-}
\end{eqnarray}
Note that they agree for $\eta < \eta'$, whereas they are complex
conjugates of one another for $\eta > \eta'$.

Of course we can only solve for the one loop corrections to the
field because we lack the higher loop contributions to the
self-energy. Suppressing spinor indices and polarities, the general
perturbative expansion takes the form,
\begin{equation}
\Xi(x) = \sum_{\ell = 0}^{\infty} \kappa^{2\ell} \Xi_{\ell}(x)
\qquad {\rm and}\,\,\, \Bigl[\Sigma\Bigr](x;x') =
\sum_{\ell=1}^{\infty} \kappa^{2\ell}
\Bigl[\Sigma_{\ell}\Bigr](x;x') \; .
\end{equation}

One substitutes these expansions into the effective Dirac equation
(\ref{Diraceqn}) and then segregates powers of $\kappa^2$,
\begin{eqnarray}
i\hspace{-.1cm}\not{\hspace{-.08cm} \partial} \Xi_0(x) & = & 0 \; , \\
\kappa^2 i\hspace{-.1cm}\not{\hspace{-.08cm} \partial} \Xi_1(x) & =
& \kappa^2 \int d^4x'
\Biggl\{\Bigl[\Sigma_1\Bigr]_{\scriptscriptstyle \!++}\!\!(x;x') +
\Bigl[\Sigma_1\Bigr]_{\scriptscriptstyle \!+-}\!\!(x;x') \Biggr\} \,
\Xi_0(x') \; , \label{source}
\end{eqnarray}
and so on. We considered the one loop correction
$\Xi_{1i}(x;\vec{k},s)$ to a spatial plane wave of helicity $s$,
\begin{equation}
\Xi_{0i}(x;\vec{k},s) = \frac{e^{-i k \eta}}{\sqrt{2k}}
u_i(\vec{k},s) e^{i \vec{k} \cdot \vec{x}} \quad {\rm where} \quad
k^{\ell} \gamma^{\ell}_{ij} u_j(\vec{k},s) = k \gamma^0_{ij}
u_j(\vec{k},s) \; .\label{freefun}
\end{equation}
In the limit of late times the source term on the right hand side of
(\ref{source}) takes the form,
\begin{eqnarray}
\kappa^2 i\hspace{-.1cm}\not{\hspace{-.08cm}\del} \Xi_1(x;\vec{k},s)
\longrightarrow \frac{\kappa^2 H^2}{16 \pi^2} \times \f{17}8 i H a
\gamma^0 \Xi_0(x;\vec{k},s) \,\, . \label{predict}
\end{eqnarray}
Hence we conclude that the late time limit of the one loop
correction to the effective mode function gives a time-dependent
enhancement of the tree order field strength \cite{MW3},
\begin{equation}
\Xi_0(x;\vec{k},s) + \kappa^2 \Xi_1(x;\vec{k},s) \longrightarrow
\Biggl\{1 \!+\! \frac{\kappa^2 H^2}{16 \pi^2} \times \frac{17}{8}
\ln(a) \Biggr\} \Xi_0(x;\vec{k},s) \,\, . \label{modefun}
\end{equation}

We must now explain how the $\comp$-number effective mode function
$\Xi(x;\vec{k},s)$ relates to the canonical fermion operator
$\Psi(x)$. Consider the perturbative expansions of the Heisenberg
operator equations for the graviton $h_{\mu\nu}(x)$ and the
(conformally rescaled) fermion $\Psi_i(x)$,
\begin{eqnarray}
h_{\mu\nu}(x) & = & h_{0\mu\nu}(x) + \kappa h_{1\mu\nu}(x) +
\kappa^2
h_{2\mu\nu}(x) + \dots \; , \\
\Psi_i(x) & = & \Psi_{0i}(x) + \kappa \Psi_{1i}(x) + \kappa^2
\Psi_{2i}(x) + \dots \; .
\end{eqnarray}
Long experience with such expansions permits us to anticipate how
the first and second order corrections to $\Psi$ depend upon the
zeroth order fields,
\begin{equation}
\Psi_1 \sim h_0 \Psi_0 \qquad , \qquad \Psi_2 \sim h_0 h_0 \Psi_0 +
\overline{\Psi}_0 \Psi_0 \Psi_0 \; .
\end{equation}

Because our state is released in free vacuum at $t=0$ ($\eta =
-1/H$), it makes sense to express the zeroth order solutions in
terms of the creation and annihilation operators of this free state,
\begin{eqnarray}
h_{0 \mu\nu}(x) & = & \int \frac{d^{D-1}k}{(2\pi)^{D-1}}
\sum_{\lambda} \Bigl\{ \epsilon_{\mu\nu}(\eta;\vec{k},\lambda) e^{i
\vec{k} \cdot \vec{x}}
\alpha(\vec{k},\lambda) \nonumber \\
& & \hspace{5cm} + \epsilon^*_{\mu\nu}(\eta;\vec{k},\lambda) e^{-i
\vec{k} \cdot \vec{x}} \alpha^{\dagger}(\vec{k},\lambda) \Bigr\} \;
, \qquad
\label{h0} \\
\Psi_{0i}(x) & = & \int \frac{d^{D-1}k}{(2\pi)^{D-1}} \sum_s \Bigl\{
\frac{e^{-ik \eta}}{\sqrt{2 k}} u_i(\vec{k},s) e^{i \vec{k}
\cdot \vec{x}} b(\vec{k},s) \nonumber \\
& & \hspace{5cm} + \frac{e^{ik \eta}}{\sqrt{2 k}}
v_i(\vec{k},\lambda) e^{-i \vec{k} \cdot \vec{x}}
c^{\dagger}(\vec{k},s) \Bigr\} \; . \qquad \label{psi0}
\end{eqnarray}
The graviton mode functions are proportional to Hankel functions
whose precise specification we do not require. The Dirac wave
functions $u_i(\vec{k}, s)$ and $v_i(\vec{k},s)$ are precisely those
of flat space by virtue of the conformal invariance of massless
fermions. The canonically normalized creation and annihilation
operators obey,
\begin{eqnarray}
\Bigl[\alpha(\vec{k},\lambda),
\alpha^{\dagger}(\vec{k}',\lambda')\Bigr] & = & \delta_{\lambda
\lambda'} (2\pi)^{D-1} \delta^{D-1}\!(\vec{k} \!-\! \vec{k}')
\label{alpop} \; , \\
\Bigl\{b(\vec{k},s), b^{\dagger}(\vec{k}',s')\Bigr\} & = & \delta_{s
s'} (2\pi)^{D-1} \delta^{D-1}\!(\vec{k} \!-\! \vec{k}') =
\Bigl\{c(\vec{k},s), c^{\dagger}(\vec{k}',s')\Bigr\} \; . \qquad
\label{bcop}
\end{eqnarray}

We can get the $\comp$-number mode function $\Xi_i(x;\vec{k},s)$
from the zeroth order field $\Psi_{0i}(x)$ by anti-commuting with
the fermion creation operator,
\begin{equation}
\Xi_{0i}(x;\vec{k},s) =
\Bigl\{\Psi_{0i}(x),b^{\dagger}(\vec{k},s)\Bigr\} = \frac{e^{-i k
\eta}}{\sqrt{2k}} u_i(\vec{k},s) e^{i \vec{k} \cdot \vec{x}} \;
.\label{Xi0}
\end{equation}
The higher order contributions to $\Psi_i(x)$ are no longer linear
in the creation and annihilation operators, so anti-commuting the
full solution $\Psi_i(x)$ with $b^{\dagger}(\vec{k},s)$ produces an
operator whose general form is,
\begin{equation}
\Bigl\{ \Psi,b^{\dagger} \Bigr\} \sim \Xi_0 + \kappa h_0 \Xi_0 +
\kappa^2 h_0 h_0 \Xi_0 + \kappa^2 \overline{\Psi}_0 \Psi_0 \Xi_0 +
O(\kappa^3) \; .
\end{equation}
The quantum-corrected fermion mode function we obtain by solving
(\ref{Diraceqn}) is the expectation value of this operator in the
presence of the state which is free vacuum at $t=0$,
\begin{equation}
\Xi_i(x;\vec{k},s) = \Bigl\langle \Omega \Bigl\vert
\Bigl\{\Psi_i(x), b^{\dagger}(\vec{k},s) \Bigr\} \Bigr\vert \Omega
\Bigr\rangle \; . \label{SKop}
\end{equation}
This is the promised relation between solving for the effective mode
function and canonical operators \cite{MW3}.

Because we have a prediction (\ref{predict}) for the late time limit
of $i \hspace{-.1cm}\not{\hspace{-.10cm} \mathcal{\partial}}
\Xi(x;\vec{k},s)$ it makes sense to act the free kinetic operator on
(\ref{SKop}),
\begin{equation}
i \hspace{-.1cm}\not{\hspace{-.10cm} \mathcal{\partial}}
\Xi(x;\vec{k},s) = \Bigl\langle \Omega \Bigl\vert \Bigl\{ i
\hspace{-.1cm}\not{\hspace{-.10cm} \mathcal{\partial}} \Psi(x),
b^{\dagger}(\vec{k},s) \Bigr\} \Bigr\vert \Omega \Bigr\rangle \; .
\label{Skop1}
\end{equation}
Of course this equation must hold order-by-order in the $\kappa$
expansions of $\Xi(x;\vec{k},s)$ and $\Psi(x)$. The order $\kappa^0$
terms vanish identically. There is no order $\kappa^1$ correction to
$\Xi(x;\vec{k},s)$, and the order $\kappa^1$ correction to $\Psi(x)$
vanishes when the expectation value is taken. The key relation for
this paper comes from taking the late time limit at order
$\kappa^2$,
\begin{equation}
\kappa^2 \Bigl\langle \Omega \Bigl\vert \Bigl\{ i \hspace{-.1cm}
\not{\hspace{-.10cm} \mathcal{\partial}} \Psi_2(x),
b^{\dagger}(\vec{k},s) \Bigr\} \Bigr\vert \Omega \Bigr\rangle
\longrightarrow \frac{\kappa^2 H^2}{ 16 \pi^2} \times \f{17}8 i H a
\gamma^0 \Xi_0(x;\vec{k},s) \; . \label{keyeqn}
\end{equation}

\section{Perturbative Operator Solution}
The purpose of this section is to work out the canonical operator
contributions to the left hand side of expression (\ref{keyeqn}). We
begin by giving the invariant action and fixing the gauge. This
defines the fermion and gravtion propagators which, in turn, give
the retarded Green's functions. We then perturbatively solve the
Heisenberg operator equations to the required order in powers of the
free fields (\ref{h0}-\ref{psi0}). Our result for $\kappa^2
i\hspace{-0.1cm}\not{\hspace{-.1cm}}\partial\Psi_2$ is reported in
Table~\ref{idelPsi}. We also report the contribution of each term to
$\kappa^2 \langle \Omega \vert
\{i\hspace{-0.1cm}\not{\hspace{-.1cm}}
\partial \Psi_2(x),b^{\dagger}(\vec{k},s)\} \vert \Omega \rangle$ in
Table~\ref{contribs}. All the analysis of this section is done in
$D$ dimensions so that ultraviolet divergences are dimensionally
regulated.

The invariant Lagrangian density of Dirac + Einstein is,
\begin{equation}
\mathcal{L} = \frac1{16 \pi G} \Bigl(R - (D\!-\!1)(D\!-\!2)
H^2\Bigr) \sqrt{-g} + \overline{\psi} e^{\mu}_{~b} \gamma^b
\Bigl(i\partial_{\mu} - \frac12 A_{\mu cd} J^{cd}\Bigr) \psi
\sqrt{-g} \; .
\end{equation}
Here $G$ is Newton's constant and $H$ is the Hubble constant. The
vierbein field is $e_{\mu b}$ and $g_{\mu\nu} \equiv e_{\mu b}
e_{\nu c} \eta^{bc}$ is the metric. The metric and
vierbein-compatible connections are,
\begin{equation}
\Gamma^{\rho}_{~\mu\nu} \equiv \frac12 g^{\rho\sigma}
\Bigl(g_{\sigma \mu , \nu} + g_{\nu \sigma , \mu} - g_{\mu \nu ,
\sigma}\Bigr) \qquad {\rm and} \qquad A_{\mu cd} \equiv e^{\nu}_{~c}
\Bigl( e_{\nu d , \mu} - \Gamma^{\rho}_{~\mu\nu} e_{\rho d}\Bigr) \;
.
\end{equation}
The Ricci scalar is,
\begin{equation}
R \equiv g^{\mu\nu} \Bigl( \Gamma^{\rho}_{~\nu\mu , \rho} -
\Gamma^{\rho}_{ ~\rho \mu , \nu} + \Gamma^{\rho}_{~\rho \sigma}
\Gamma^{\sigma}_{~\nu\mu} - \Gamma^{\rho}_{~\nu \sigma}
\Gamma^{\sigma}_{~\rho \mu} \Bigr) \; .
\end{equation}
The gamma matrices $\gamma^b_{ij}$ have spinor indices $i, j \in
\{1,2,3,4\}$, obey the usual anti-commutation relations and give the
usual Lorentz generators,
\begin{equation}
\{\gamma^b , \gamma^c\} = -2 \eta^{bc} I \qquad , \qquad J^{bc}
\equiv \frac{i}4 [\gamma^b ,\gamma^c] \; .
\end{equation}

It is useful to conformally rescale the vierbein by the de Sitter
scale factor $a(t)$,
\begin{equation}
e_{\beta b} \equiv a \, \widetilde{e}_{\beta b} \qquad
\Longrightarrow \qquad e^{\beta b} = a^{-1} \, \widetilde{e}^{\beta
b} \; .
\end{equation}
Of course this implies a rescaled metric $\widetilde{g}_{\mu\nu}$,
\begin{equation}
g_{\mu\nu} = a^2 \, \widetilde{g}_{\mu\nu} \equiv a^2 \Bigl(
\eta_{\mu\nu} + \kappa h_{\mu\nu}(x)\Bigr) \quad {\rm where} \quad a
= -\frac1{H \eta} = e^{H t} \; .\label{confg}
\end{equation}
The old connections can be expressed as follows in terms of the ones
formed from the rescaled fields,
\begin{eqnarray}
\Gamma^{\rho}_{~\mu\nu} & = & a^{-1} \Bigl(\delta^{\rho}_{~\mu} \,
a_{,\nu} \!+\!  \delta^{\rho}_{~\nu} \, a_{,\mu} \!-\!
\widetilde{g}^{\rho\sigma} \, a_{,\sigma} \,
\widetilde{g}_{\mu\nu}\Bigr) + \widetilde{\Gamma}^{\rho}_{
~\mu\nu} \label{confG} \; \\
A_{\mu cd} & = &-a^{-1} \Bigl(\widetilde{e}^{\nu}_{~c} \,
\widetilde{e}_{\mu d} \!-\!  \widetilde{e}^{\nu}_{~d} \,
\widetilde{e}_{\mu c} \Bigr) a_{,\nu} + \widetilde{A}_{\mu cd} \; .
\end{eqnarray}
We define rescaled fermion fields as,
\begin{equation}
\Psi \equiv a^{\frac{D-1}2} \, \psi \qquad {\rm and} \qquad
\overline{\Psi} \equiv a^{\frac{D-1}2} \, \overline{\psi} \; .
\end{equation}

We employ Lorentz symmetric gauge, $e_{\mu b} = e_{b \mu}$, which
permits one to perturbatively determine the vierbein in terms of the
metric and their respective backgrounds \cite{RPW2},
\begin{equation}
\widetilde{e}[\widetilde{g}]_{\beta b} \equiv
\Bigl(\sqrt{\widetilde{g} \eta^{-1}} \, \Bigr)_{\!\beta}^{~\gamma}
\, \eta_{\gamma b} = \eta_{\beta b} + \frac12 \kappa h_{\beta b} -
\frac18 \kappa^2 h_{\beta}^{~\gamma} h_{\gamma b} + \dots
\end{equation}
Here and throughout this paper graviton indices are raised and
lowered with the Lorentz metric, e.g., $h^{\mu\nu} = \eta^{\mu\rho}
\eta^{\nu\sigma} h_{\rho\sigma}$. The same convention applies as
well to derivatives ($\partial^{\mu} = \eta^{\mu\nu}
\partial_{\nu}$) and gamma matrices
($\gamma_{\mu} = \eta_{\mu\nu} \gamma^{\nu}$). The general
coordinate freedom is fixed by adding the gauge fixing term,
\begin{equation}
\Delta \mathcal{L} = -\frac12 a^{D-2} \eta^{\mu\nu} F_{\mu} F_{\nu}
\qquad F_{\mu} \equiv \eta^{\rho\sigma} \Bigl( h_{\mu\rho , \sigma}
\!-\! \frac12 h_{\rho \sigma , \mu} \!+\! (D\!-\!2) H a h_{\mu \rho}
\delta^0_{\sigma}\Bigr) . \label{ourgauge}
\end{equation}
After some judicious partial integrations the gauge fixed Lagrangian
density has the following expansion,
\begin{eqnarray}
\lefteqn{\mathcal{L}_{\scriptscriptstyle {\rm GF}} = \overline{\Psi}
i \hspace{-.1cm}\not{\hspace{-.1cm} \partial} \Psi + \frac{\kappa}2
\Bigl[h \overline{\Psi} i \hspace{-.1cm}\not{\hspace{-.1cm}
\partial} \Psi \!-\!
h^{\mu\nu} \overline{\Psi} \gamma_{\mu} i \partial_{\nu} \Psi \!-\!
h_{\mu\rho , \sigma} \overline{\Psi} \gamma^{\mu} J^{\rho\sigma}
\Psi
\Bigr] } \nonumber \\
& & + \kappa^2 \Bigl[\frac18 h^2 \!-\! \frac14 h^{\rho\sigma}
h_{\rho\sigma}\Bigr] \overline{\Psi} i
\hspace{-.1cm}\not{\hspace{-.1cm}
\partial} \Psi \!+\! \kappa^2 \Bigl[-\frac14 h h^{\mu\nu} \!+\! \frac38
h^{\mu\rho} h_{\rho}^{~\nu}\Bigr] \overline{\Psi} \gamma_{\mu} i
\partial_{\nu}
\Psi \nonumber \\
& & + \kappa^2 \Bigl[-\frac14 h h_{\mu \rho , \sigma} \!+\! \frac18
h^{\nu}_{~\rho} h_{\nu \sigma , \mu} \!+\! \frac14 (h^{\nu}_{~\mu}
h_{\nu\rho})_{,\sigma} \!+\! \frac14 h^{\nu}_{~\sigma} h_{\mu\rho
,\nu}\Bigr]
\overline{\Psi} \gamma^{\mu} J^{\rho\sigma} \Psi + O(\kappa^3) \nonumber \\
& & \hspace{3.5cm} + \frac12 h^{\mu\nu} D_{\mu\nu}^{~~\rho \sigma}
h_{\rho\sigma} + \Bigl({\rm Pure\ Gravity\ Interactions}\Bigr) \; .
\qquad \label{Dexp}
\end{eqnarray}
The explicit form of the graviton kinetic operator
$D_{\mu\nu}^{~~\rho\sigma}$ is not needed here; it can be found in
ref.~\cite{MW2}.

The $++$ and $+-$ fermion propagators are related to the conformal
scalar propagator in the usual way,
\begin{equation}
i [S]_{\scriptscriptstyle \!+\pm}(x \!-\! x') = i\hspace{-.1cm}
\not{\hspace{-.1cm} \partial} \, i\Delta^{\rm
cf}_{\scriptscriptstyle \!+\pm}\! (x \!-\! x') \equiv
i\hspace{-.1cm} \not{\hspace{-.1cm} \partial} \times
\frac{\Gamma(\frac{D}2 \!-\! 1)}{4 \pi^{\frac{D}2}} \, \frac1{\Delta
x^{D-2}_{\scriptscriptstyle +\pm}} \; . \label{cf}
\end{equation}
The two conformal coordinate intervals $\Delta
x^2_{\scriptscriptstyle +\pm}$ were defined in
(\ref{x++}-\ref{x+-}).

The graviton propagator takes the form of a sum of three scalar
propagators times constant tensor factors \cite{TW2RPW0},
\begin{equation}
i\Bigl[\mbox{}_{\mu\nu}
\Delta_{\rho\sigma}\Bigr]_{\scriptscriptstyle \!+\pm} \!(x;x') =
\sum_{I=A,B,C} \Bigl[\mbox{}_{\mu\nu} T_{\rho\sigma}^{I}\Bigr] \,
i\Delta^I_{\scriptscriptstyle \!+\pm}\!(x;x') \; . \label{gravprop}
\end{equation}
Because our gauge (\ref{ourgauge}) treats time and space differently
it is useful to have expressions to the purely spatial parts of the
Lorentz metric and the Kronecker delta,
\begin{equation}
\overline{\eta}_{\mu\nu} \equiv \eta_{\mu\nu} + \delta^0_{\mu}
\delta^0_{\nu} \qquad {\rm and} \qquad \overline{\delta}^{\mu}_{\nu}
\equiv \delta^{\mu}_{\nu} - \delta_0^{\mu} \delta^0_{\nu} \; .
\end{equation}
With this convention, the three tensor factors in (\ref{gravprop})
are,
\begin{eqnarray}
\Bigl[{}_{\mu\nu} T^A_{\rho\sigma}\Bigr] & = & 2 \,
\overline{\eta}_{\mu (\rho} \overline{\eta}_{\sigma) \nu} -
\frac2{D\!-\! 3} \overline{\eta}_{\mu\nu}
\overline{\eta}_{\rho \sigma} \; , \label{Atensor} \\
\Bigl[{}_{\mu\nu} T^B_{\rho\sigma}\Bigr] & = & -4 \delta^0_{(\mu}
\overline{\eta}_{\nu) (\rho} \delta^0_{\sigma)} \; , \\
\Bigl[{}_{\mu\nu} T^C_{\rho\sigma}\Bigr] & = & \frac2{(D \!-\!2) (D
\!-\!3)} \Bigl[(D \!-\!3) \delta^0_{\mu} \delta^0_{\nu} +
\overline{\eta}_{\mu\nu}\Bigr] \Bigl[(D \!-\!3) \delta^0_{\rho}
\delta^0_{\sigma} + \overline{\eta}_{\rho \sigma}\Bigr] \; .
\end{eqnarray}
We follow the usual convention that parenthesized indices are
symmetrized.

The three scalar propagators in (\ref{gravprop}) can be expressed in
terms of the appropriate de Sitter invariant length function
$y_{\scriptscriptstyle +\pm}\!(x;x')$,
\begin{equation}
y_{\scriptscriptstyle +\pm}\!(x;x') \equiv a(t) a(t') H^2 \Delta
x^2_{\scriptscriptstyle \!+\pm}(x;x') \; .
\end{equation}
The $B$-type and $C$ type propagators are hypergeometric functions,
\begin{eqnarray}
i\Delta^B_{\scriptscriptstyle \!+\pm}\!(x;x') & = &
\frac{H^{D-2}}{(4\pi)^{ \frac{D}2}} \frac{\Gamma(D\!-\!2)
\Gamma(1)}{\Gamma(\frac{D}2)} \,
\mbox{}_2F_1\Bigl(D\!-\!2,1;\frac{D}2;1 \!-\! \frac{y_{+\pm}}4\Bigr)
\; ,
\label{FDB} \\
i\Delta^C_{\scriptscriptstyle \!+\pm}(x;x') & = &
\frac{H^{D-2}}{(4\pi)^{ \frac{D}2}} \frac{\Gamma(D\!-\!3)
\Gamma(2)}{\Gamma(\frac{D}2)} \,
\mbox{}_2F_1\Bigl(D\!-\!3,2;\frac{D}2;1 \!-\! \frac{y_{+\pm}}4\Bigr)
\; . \label{FDC}
\end{eqnarray}
The $A$-type propagator has the intimidating expansion,
\begin{eqnarray}
\lefteqn{i \Delta^A_{\scriptscriptstyle \!+\pm}\!(x;x') =  i
\Delta^{\rm
cf}_{\scriptscriptstyle \!+\pm}\!(x;x') } \nonumber \\
& & \hspace{-.7cm} + \frac{H^{D-2}}{(4\pi)^{\frac{D}2}}
\frac{\Gamma(D \!-\! 1)}{\Gamma(\frac{D}2)} \left\{\!
\frac{D}{D\!-\! 4} \frac{\Gamma^2(\frac{D}2)}{\Gamma(D \!-\! 1)}
\Bigl(\frac4{y_{+\pm}}\Bigr)^{\frac{D}2 -2} \!\!\!\!\!\! - \pi
\cot\Bigl(\frac{\pi}2 D\Bigr) + \ln(a a') \!\right\} \nonumber \\
& & \hspace{-.7cm} + \frac{H^{D-2}}{(4\pi)^{\frac{D}2}} \!
\sum_{n=1}^{\infty}\! \left\{\!\frac1{n} \frac{\Gamma(n \!+\! D
\!-\! 1)}{ \Gamma(n \!+\! \frac{D}2)} \Bigl(\frac{y_{+\pm}}4
\Bigr)^n \!\!\!\! - \! \frac1{n \!-\! \frac{D}2 \!+\! 2}
\frac{\Gamma(n \!+\!  \frac{D}2 \!+\! 1)}{ \Gamma(n \!+\! 2)}
\Bigl(\frac{y_{+\pm}}4 \Bigr)^{n - \frac{D}2 +2} \!\right\} \! .
\qquad \label{DeltaA}
\end{eqnarray}

We need retarded Green's functions in order to develop an expansion
for the full fields in terms of the free fields of the initial time.
There is a very simple relation between the retarded Green's
function of any field and the corresponding $++$ and $+-$
propagators. If the field's kinetic operator is $\mathcal{D}$ then
the two propagators obey,
\begin{equation}
\mathcal{D} \, i\Delta_{\scriptscriptstyle \!++}\!(x;x') = i
\delta^D\!(x\!-\!x') \qquad {\rm and} \qquad \mathcal{D} \,
i\Delta_{\scriptscriptstyle \!+-}\!(x;x') = 0 \; . \label{propeqns}
\end{equation}
The associated retarded Green's function is,
\begin{equation}
G(x;x') = i \Bigl[i\Delta_{\scriptscriptstyle \!++}\!(x;x') - i
\Delta_{\scriptscriptstyle \!+-}\!(x;x')\Bigr] \; . \label{Gret}
\end{equation}
From (\ref{propeqns}) one easily sees that it obeys the required
equation,
\begin{equation}
\mathcal{D} \, G(x;x') = -\delta^D\!( x\!-\!x') \; .
\end{equation}
It also obeys the retarded condition of vanishing for $\eta < \eta'$
because the conformal coordinate intervals (\ref{x++}) and
(\ref{x+-}) are equal in that case.

\begin{table}

\vbox{\tabskip=0pt \offinterlineskip
\def\tablerule{\noalign{\hrule}}
\halign to390pt {\strut#& \vrule#\tabskip=1em plus2em& \hfil#\hfil&
\vrule#& \hfil#\hfil& \vrule#\tabskip=0pt\cr \tablerule
\omit&height4pt&\omit&&\omit&\cr &&$\!\!\!\!{\rm Term}\!\!\!\!$ &&
$\!\!\!\! {\rm Contribution\ to} \; \kappa^2 i \hspace{-.1cm}
\not{\hspace{-.1cm} \partial} \Psi_2(x) \!\!\!\!$ & \cr
\omit&height4pt&\omit&&\omit&\cr \tablerule
\omit&height2pt&\omit&&\omit&\cr && 1a && $-\frac14 \kappa^2
h_0^{~\mu\nu}(x) \gamma_{\mu} i \partial_{\nu} i \hspace{-.1cm}
\not{\hspace{-.1cm} \partial}\int \! d^Dx' \, G_{\rm cf}(x\!-\!x')
h_0^{\rho\sigma}(x') \gamma_{\rho} i \partial'_{\sigma} \Psi_0(x')$
& \cr \omit&height2pt&\omit&&\omit&\cr \tablerule
\omit&height2pt&\omit&&\omit&\cr && 1b && $-\frac14 \kappa^2
h_0^{~\mu\nu}(x) \gamma_{\mu} i \partial_{\nu} i \hspace{-.1cm}
\not{\hspace{-.1cm} \partial}\int \! d^Dx' \, G_{\rm cf}(x\!-\!x')
h_0^{\rho\sigma , \beta}(x') \gamma_{\rho} J_{\sigma\beta}
\Psi_0(x')$ & \cr \omit&height2pt&\omit&&\omit&\cr \tablerule
\omit&height2pt&\omit&&\omit&\cr && 2a && $-\frac14 \kappa^2
h_0^{~\mu\nu ,\alpha}(x) \gamma_{\mu} J_{\nu\alpha} i \hspace{-.1cm}
\not{\hspace{-.1cm} \partial}\int \! d^Dx' \, G_{\rm cf}(x\!-\!x')
h_0^{\rho\sigma}(x') \gamma_{\rho} i
\partial'_{\sigma} \Psi_0(x')$ & \cr
\omit&height2pt&\omit&&\omit&\cr \tablerule
\omit&height2pt&\omit&&\omit&\cr && 2b && $-\frac14 \kappa^2
h_0^{~\mu\nu ,\alpha}(x) \gamma_{\mu} J_{\nu\alpha} i \hspace{-.1cm}
\not{\hspace{-.1cm} \partial}\int \! d^Dx' \, G_{\rm cf}(x\!-\!x')
h_0^{\rho\sigma ,\beta}(x') \gamma_{\rho} J_{\sigma\beta}
\Psi_0(x')$ & \cr \omit&height2pt&\omit&&\omit&\cr \tablerule
\omit&height2pt&\omit&&\omit&\cr && 3a && $-\frac14 \kappa^2 \int \!
d^Dx' \, [\mbox{}^{\mu\nu} G^{\rho\sigma}](x;x')
\overline{\Psi}_0(x') \gamma_{\rho} i \partial_{\sigma}' \Psi_0(x')
\times \gamma_{\mu} i \partial_{\nu} \Psi_0(x)$ & \cr
\omit&height2pt&\omit&&\omit&\cr \tablerule
\omit&height2pt&\omit&&\omit&\cr && 3b && $\frac14 \kappa^2 \int \!
d^Dx' \, [\mbox{}^{\mu\nu} G^{\rho\sigma}](x;x')
[\overline{\Psi}_0(x') \gamma_{\rho} J_{\sigma\beta}
\Psi_0(x')]^{,\beta} \times \gamma_{\mu} i \partial_{\nu} \Psi_0(x)$
& \cr \omit&height2pt&\omit&&\omit&\cr \tablerule
\omit&height2pt&\omit&&\omit&\cr && 4a && $-\frac14 \kappa^2
\partial^{\alpha} \! \int \! d^Dx' \,
[\mbox{}^{\mu\nu} G^{\rho\sigma}](x;x') \overline{\Psi}_0(x')
\gamma_{\rho} i
\partial_{\sigma}' \Psi_0(x') \times \gamma_{\mu} J_{\nu\alpha}
\Psi_0(x)$ & \cr \omit&height2pt&\omit&&\omit&\cr \tablerule
\omit&height2pt&\omit&&\omit&\cr && 4b && $\!\!\frac14 \kappa^2
\partial^{\alpha} \! \int \! d^Dx' \,
[\mbox{}^{\mu\nu} G^{\rho\sigma}](x;x') [\overline{\Psi}_0(x')
\gamma_{\rho} J_{\sigma\beta} \Psi_0(x')]^{,\beta} \times
\gamma_{\mu} J_{\nu\alpha} \Psi_0(x) \!\!$ & \cr
\omit&height2pt&\omit&&\omit&\cr \tablerule
\omit&height2pt&\omit&&\omit&\cr && 5 && $-\frac38 \kappa^2
h_0^{~\mu\nu}(x) \, h_0^{\rho\sigma}(x) \, \eta_{\mu\rho}
\gamma_{\nu} i \partial_{\sigma} \Psi_0(x)$ & \cr
\omit&height2pt&\omit&&\omit&\cr \tablerule
\omit&height2pt&\omit&&\omit&\cr && 6 && $-\frac18 \kappa^2
h_0^{\mu\nu}(x) \, h_0^{\rho\sigma , \alpha}(x) \, \eta_{\mu\rho}
\gamma_{\alpha} J_{\nu\sigma} \Psi_0(x)$ & \cr
\omit&height2pt&\omit&&\omit&\cr \tablerule
\omit&height2pt&\omit&&\omit&\cr && 7 && $-\frac14 \kappa^2
[h_0^{\mu\nu}(x) \, h_0^{\rho\sigma}(x)]^{, \alpha} \,
\eta_{\mu\rho} \gamma_{\nu} J_{\sigma\alpha} \Psi_0(x)$ & \cr
\omit&height2pt&\omit&&\omit&\cr \tablerule
\omit&height2pt&\omit&&\omit&\cr && 8 && $-\frac14 \kappa^2
h_0^{\mu\nu}(x) \, h_0^{\rho\sigma , \alpha}(x) \, \eta_{\mu\alpha}
\gamma_{\rho} J_{\sigma\nu} \Psi_0(x)$ & \cr
\omit&height2pt&\omit&&\omit&\cr \tablerule
\omit&height2pt&\omit&&\omit&\cr}}

\caption{Free Field Expansion of $\kappa^2 i \hspace{-.1cm}
\not{\hspace{-.1cm}
\partial} \Psi_2(x)$}
\label{idelPsi}

\end{table}

The Heisenberg operator equation for the fermion is,
\begin{eqnarray}
\lefteqn{i \hspace{-.1cm}\not{ \hspace{-.1cm} \partial} \Psi =
\frac{\kappa}2 \Bigl\{ -h i \hspace{-.1cm}\not{\hspace{-.1cm}
\partial} \!+\! h^{\mu\nu} \gamma_{\mu} i \partial_{\nu} \!+\!
h_{\mu\rho , \sigma} \gamma^{\mu} J^{\rho\sigma} \Bigr\} \Psi }\nonumber \\
& & \hspace{-.5cm} - \kappa^2 \Bigl\{\frac18 h^2 \!-\! \frac14
h^{\rho\sigma} h_{\rho\sigma}\Bigr] i
\hspace{-.1cm}\not{\hspace{-.1cm}
\partial} \Psi + \kappa^2 \Bigl\{\frac14 h h^{\mu\nu} \!-\! \frac38 h^{\mu\rho}
h_{\rho}^{~\nu}\Bigr]  \gamma_{\mu} i \partial_{\nu} \Psi \nonumber \\
& & \hspace{-.5cm} + \kappa^2 \Biggl\{\frac14 h h_{\mu \rho ,
\sigma} \!-\! \frac18 h^{\nu}_{~\rho} h_{\nu \sigma , \mu} \!-\!
\frac14 (h^{\nu}_{~\mu} h_{\nu\rho})_{,\sigma} \!-\! \frac14
h^{\nu}_{~ \sigma} h_{\mu\rho ,\nu}\Biggr\}  \gamma^{\mu}
J^{\rho\sigma} \Psi \!+\! O(\kappa^3) \; . \qquad \label{Diracop}
\end{eqnarray}
We only require the analogous equation for the graviton to first
order, and we only need the terms that involve fermions,
\begin{eqnarray}
\lefteqn{D_{\mu\nu}^{~~\rho \sigma} h_{\rho\sigma} = \frac{\kappa}2
\Biggl\{-\eta_{\mu\nu} \overline{\Psi} i
\hspace{-.1cm}\not{\hspace{-.1cm} \partial} \Psi \!+\!
\overline{\Psi} \gamma_{\mu} i \partial_{\nu} \Psi \!-\!
\partial^{\sigma} \Bigl[
\overline{\Psi} \gamma_{\mu} J_{\nu\sigma} \Psi\Bigr] \Biggr\}
+ O(\kappa^2) } \nonumber \\
& & \hspace{6cm} + \Bigl({\rm Pure\ Gravity\ Interactions}\Bigr) \;
. \qquad \label{grop}
\end{eqnarray}
The next step is to expand Heisenberg operators in powers of
$\kappa$,
\begin{equation}
\Psi = \Psi_0 + \kappa \Psi_1 + \kappa^2 \Psi_2 + \dots \quad ,
\quad h_{\mu\nu} = h_{0 \mu\nu} + \kappa h_{1 \mu\nu} + \dots \; .
\end{equation}
Of course the zeroth order equations ($D_{\mu\nu}^{~~\rho\sigma}
h_{0 \rho\sigma} = 0$ and $i\hspace{-.1cm} \not{\hspace{-.1cm}\del}
\Psi_0 = 0$) just give the zeroth order solutions $h_{0\mu\nu}$ and
$\Psi_0$ we already encountered in expressions (\ref{h0}) and
(\ref{psi0}), respectively. The order $\kappa$ fermion equation
implies,
\begin{equation}
i \hspace{-.1cm} \not{\hspace{-.08cm} \del} \Psi_1 = \frac12
\Bigl[-h_0 \, i\hspace{-.1cm} \not{\hspace{-.1cm} \partial} +
h_0^{\mu\nu} \gamma_{\mu} i \partial_{\nu} + h_0^{\mu\rho,\sigma}
\gamma_{\mu} J_{\rho\sigma} \Bigr] \Psi_0 = \frac12 \Bigl[
h_0^{\mu\nu} \gamma_{\mu} i \partial_{\nu} + h_0^{\mu\rho,\sigma}
\gamma_{\mu} J_{\rho\sigma} \Bigr] \Psi_0 \; .
\end{equation}
Hence the order $\kappa$ correction to the fermion operator is,
\begin{equation}
\Psi_{\!1}\!(x) = -\frac12
i\hspace{-.1cm}\not{\hspace{-.1cm}\partial} \int d^{D}x'G_{\rm
cf}(x\!-\!x')\Bigl[\,h_0^{\mu\nu}\!(x') \gamma_{\mu} i
\partial_{\nu}' + h_0^{\mu\rho,\sigma}\!(x') \gamma_{\mu} J_{\rho\sigma}
\Bigr] \Psi_{\!0}\!(x') .\label{Psi1}
\end{equation}
In the same way we obtain the first order correction to the
graviton,
\begin{eqnarray}
\lefteqn{h_1^{\mu\nu}(x) = -\frac12 \int d^{D}x' [\mbox{}^{\mu\nu}
G^{\rho\sigma}]\!(x;x') \Biggl\{\overline{\Psi}_{\!0}\!(x')
\gamma_{\rho}
i\partial_{\sigma}' \Psi_{\!0}\!(x') } \nonumber \\
& & \hspace{2.5cm} - \partial^{\prime \alpha}
\Bigl[\overline{\Psi}_{\!0}\!(x') \gamma_{\rho} J_{\sigma\alpha}
\Psi_{\!0}\!(x')\Bigr] \Biggr\} + \Bigl({\rm Pure\ Gravity\
Terms}\Bigr) \; . \qquad \label{h1}
\end{eqnarray}

This brings us to the order $\kappa^2$ correction to the fermion. We
can of course drop any factors of $i\hspace{-.15cm}
\not{\hspace{-.15cm\partial} \Psi_0} = 0$. With some further
simplifications based on the first order equations we reach the
form,
\begin{eqnarray}
\lefteqn{i \hspace{-.1cm} \not{\hspace{-.10cm} \mathcal{\partial}}
\Psi_2 = \frac12 \Bigl[ h_0^{\mu\nu} \gamma_{\mu} i \partial_{\nu} +
h_0^{\mu \rho , \sigma} \gamma_{\mu} J_{\rho\sigma} \Bigr] \Psi_1 +
\frac12 \Bigl[ h_1^{\mu\nu} \gamma_{\mu} i \partial_{\nu} +
h_1^{\mu \rho , \sigma} \gamma_{\mu} J_{\rho\sigma} \Bigr] \Psi_0 } \nonumber\\
& & \hspace{-.5cm} -\frac38 h_0^{\mu\rho} h_{0 ~\rho}^{~\nu}
\gamma_{\mu} i \partial_{\nu} \Psi_0 \!-\! \Bigl[\frac18 h_{0
~\nu}^{~\rho} h^{\nu \sigma , \mu} \!+\! \frac14 \Bigl(h_{0
~\nu}^{~\mu} h_0^{~ \nu\rho} \Bigr)^{,\sigma} \!\!+\! \frac14 h_{0
~\nu}^{~\sigma} h^{\mu \rho , \nu} \Bigr] \gamma_{\mu}
J_{\rho\sigma} \Psi_0 \; . \qquad\label{Psi2}
\end{eqnarray}
Table~\ref{idelPsi} gives the free field expansion of $i
\hspace{-.15cm} \not{\hspace{-.10cm} \mathcal{\partial}} \Psi_2$,
excepting only the contributions from the pure gravity corrections
to $h_{1\mu\nu}$ which vanish when the expectation value in
(\ref{keyeqn}) is taken.

Each contribution to Table~\ref{idelPsi} contains three free fields.
It remains to evaluate the source term (\ref{keyeqn}),
\begin{equation}
\kappa^2 \Bigl\langle \Omega \Bigl\vert \Bigl\{i\hspace{-.1cm}
\not{\hspace{-.1cm}\partial} \Psi_{2}(x), b^{+}(\vec{k},s)\Bigr\}
\Bigr\vert \Omega \Bigr\rangle + O(\kappa^4) \; . \label{Skop2}
\end{equation}
This is done by using the anti-commutator to absorb a $\Psi_0$ and
then exploiting the fundamental Schwinger-Keldysh relation
(\ref{fund}) to express the expectation value of the two remaining
free fields in terms of the propagator of appropriate polarity. To
be definite, suppose the two remaining free fields are scalars
$\varphi_0(x)$ and $\varphi_0(x')$. Here is where the factor
ordering matters. From relation (\ref{fund}) we see that the $+-$
propagator emerges from the order $\varphi_0(x') \times \varphi(x)$,
\begin{equation}
\Bigl\langle \Omega \Bigl\vert \varphi_0(x') \varphi_0(x) \Bigr\vert
\Omega \Bigr\rangle = i \Delta_{\!+-}(x;x') \; .
\end{equation}
The order $\varphi_0(x) \times \varphi_0(x')$ gives the $-+$
propagator, however, this is equivalent to the $++$ propagator when
account is taken of the factor of $\theta(\eta \!-\! \eta')$ in the
retarded Green's function that is always present,
\begin{eqnarray}
G(x;x') \times \Bigl\langle \Omega \Bigl\vert \varphi_0(x)
\varphi_0(x') \Bigr\vert \Omega \Bigr\rangle & = & G(x;x') \times i
\Delta_{\!-+}(x;x')
\; , \\
& = & G(x;x') \times i\Delta_{\!++}(x;x') \; .
\end{eqnarray}

\begin{table}

\vbox{\tabskip=0pt \offinterlineskip
\def\tablerule{\noalign{\hrule}}
\halign to390pt {\strut#& \vrule#\tabskip=1em plus2em& \hfil#\hfil&
\vrule#& \hfil#\hfil& \vrule#\tabskip=0pt\cr \tablerule
\omit&height4pt&\omit&&\omit&\cr &&$\!\!\!\!{\rm Term}\!\!\!\!$ &&
$\!\!\!\! {\rm Contribution\ to} \; \kappa^2 \Bigl\langle \Omega
\Bigl\vert \Bigl\{ i \hspace{-.1cm} \not{\hspace{-.1cm} \partial}
\Psi_2(x), b^{\dagger}(\vec{k},s)\Bigr\} \Bigr\vert \Omega
\Bigr\rangle \!\!\!\!$ & \cr \omit&height4pt&\omit&&\omit&\cr
\tablerule \omit&height2pt&\omit&&\omit&\cr && 1a &&
$\frac{i\kappa^2}4 \int \! d^Dx' \, i[\mbox{}^{\mu\nu}
\Delta^{\rho\sigma}]_{++}\!(x;x') \, \gamma_{\mu} \partial_{\nu}
\hspace{-.1cm} \not{\hspace{-.1cm} \partial} \, G_{\rm cf}(x\!-\!x')
\gamma_{\rho} \partial'_{\sigma} \Xi_0(x')$ & \cr
\omit&height2pt&\omit&&\omit&\cr \tablerule
\omit&height2pt&\omit&&\omit&\cr && 1b && $\frac{\kappa^2}4 \int \!
d^Dx' \, \partial^{\prime \beta} i[\mbox{}^{\mu\nu}
\Delta^{\rho\sigma}]_{++}\!(x;x') \, \gamma_{\mu}
\partial_{\nu} \hspace{-.1cm} \not{\hspace{-.1cm} \partial} \,
G_{\rm cf}(x\!-\!x') \gamma_{\rho} J_{\sigma\beta} \Xi_0(x')$ & \cr
\omit&height2pt&\omit&&\omit&\cr \tablerule
\omit&height2pt&\omit&&\omit&\cr && 2a && $\frac{\kappa^2}4 \int \!
d^Dx' \, \partial^{\alpha} i[\mbox{}^{\mu\nu}
\Delta^{\rho\sigma}]_{++}\!(x;x') \, \gamma_{\mu} J_{\nu\alpha}
\hspace{-.1cm} \not{\hspace{-.1cm} \partial} \, G_{\rm cf}(x\!-\!x')
\gamma_{\rho} \partial'_{\sigma} \Xi_0(x')$ & \cr
\omit&height2pt&\omit&&\omit&\cr \tablerule
\omit&height2pt&\omit&&\omit&\cr && 2b && $-\frac{i \kappa^2}4 \int
\! d^Dx' \, \partial^{\alpha}
\partial^{\prime \beta} i[\mbox{}^{\mu\nu} \Delta^{\rho\sigma}]_{++}\!(x;x') \,
\gamma_{\mu} J_{\nu\alpha} \hspace{-.1cm} \not{\hspace{-.1cm}
\partial} \,
G_{\rm cf}(x\!-\!x') \gamma_{\rho} J_{\sigma\beta} \Xi_0(x')$ & \cr
\omit&height2pt&\omit&&\omit&\cr \tablerule
\omit&height2pt&\omit&&\omit&\cr && 3a && $\frac{i \kappa^2}4 \int
\! d^Dx' \, [\mbox{}^{\mu\nu} G^{\rho\sigma}](x;x') \, \gamma_{\mu}
\partial_{\nu}
\hspace{-.1cm} \not{\hspace{-.1cm} \partial} \, i\Delta^{\rm
cf}_{+-}(x\!-\!x') \gamma_{\rho} \partial_{\sigma}' \Xi_0(x')$ & \cr
\omit&height2pt&\omit&&\omit&\cr \tablerule
\omit&height2pt&\omit&&\omit&\cr && 3b && $\frac{\kappa^2}4 \int \!
d^Dx' \, \partial^{\prime \beta} [\mbox{}^{\mu\nu}
G^{\rho\sigma}](x;x') \, \gamma_{\mu}
\partial_{\nu}
\hspace{-.1cm} \not{\hspace{-.1cm} \partial} \, i\Delta^{\rm
cf}_{+-}(x\!-\!x') \gamma_{\rho} J_{\sigma\beta} \Xi_0(x')$ & \cr
\omit&height2pt&\omit&&\omit&\cr \tablerule
\omit&height2pt&\omit&&\omit&\cr && 4a && $\frac{\kappa^2}4 \! \int
\! d^Dx' \, \partial^{\alpha} [\mbox{}^{\mu\nu}
G^{\rho\sigma}](x;x') \, \gamma_{\mu} J_{\nu\alpha} \hspace{-.1cm}
\not{\hspace{-.1cm} \partial} \, i\Delta^{\rm cf}_{+-}(x\!-\!x')
\gamma_{\rho} \partial_{\sigma}' \Xi_0(x')$ & \cr
\omit&height2pt&\omit&&\omit&\cr \tablerule
\omit&height2pt&\omit&&\omit&\cr && 4b && $\!\!-\frac{i \kappa^2}4
\! \int \! d^Dx' \,
\partial^{\alpha}
\partial^{\prime \beta} [\mbox{}^{\mu\nu} G^{\rho\sigma}](x;x')
\gamma_{\mu} J_{\nu\alpha} \hspace{-.1cm} \not{\hspace{-.1cm}
\partial} \,
i\Delta^{\rm cf}_{+-}(x\!-\!x') \gamma_{\rho} J_{\sigma\beta}
\Xi_0(x')\!\!$ & \cr \omit&height2pt&\omit&&\omit&\cr \tablerule
\omit&height2pt&\omit&&\omit&\cr && 5 && $-\frac{3 i \kappa^2}8
i[\mbox{}^{\mu\nu} \Delta^{\rho\sigma}](x;x) \, \eta_{\mu\rho}
\gamma_{\nu} \partial_{\sigma} \Xi_0(x)$ & \cr
\omit&height2pt&\omit&&\omit&\cr \tablerule
\omit&height2pt&\omit&&\omit&\cr && 6 && $-\frac{\kappa^2}8 \lim_{x'
\rightarrow x} \partial^{\prime \alpha} i[\mbox{}^{\mu\nu}
\Delta^{\rho\sigma}](x;x') \, \eta_{\mu\rho} \gamma_{\alpha}
J_{\nu\sigma} \Xi_0(x)$ & \cr \omit&height2pt&\omit&&\omit&\cr
\tablerule \omit&height2pt&\omit&&\omit&\cr && 7 &&
$-\frac{\kappa^2}4 \partial^{\alpha} i[\mbox{}^{\mu\nu}
\Delta^{\rho\sigma}](x;x) \, \eta_{\mu\rho} \gamma_{\nu}
J_{\sigma\alpha} \Xi_0(x)$ & \cr \omit&height2pt&\omit&&\omit&\cr
\tablerule \omit&height2pt&\omit&&\omit&\cr && 8 &&
$-\frac{\kappa^2}4 \lim_{x' \rightarrow x} \partial^{\prime \alpha}
i[\mbox{}^{\mu\nu} \Delta^{\rho\sigma}](x;x') \, \eta_{\mu\alpha}
\gamma_{\rho} J_{\sigma\nu} \Xi_0(x)$ & \cr
\omit&height2pt&\omit&&\omit&\cr \tablerule
\omit&height2pt&\omit&&\omit&\cr}}

\caption{Contribution to $\kappa^2 \Bigl\langle \Omega \Bigl\vert
\Bigl\{i \hspace{-.15cm} \not{\hspace{-.10cm} \partial}
\Psi_2(x),b^{\dagger}(\vec{k},s)\Bigr\} \Bigr\vert \Omega
\Bigr\rangle$ from each term in the free field expansion.}
\label{contribs}
\end{table}

As an example we work out the $(4b)$ term. It is useful to begin by
partially integrating the $\partial^{\prime \beta}$ without
retaining the temporal surface term,
\begin{equation}
(4b) \longrightarrow -\frac{\kappa^2}4 \int d^Dx' \,
\partial^{\alpha}
\partial^{\prime \beta} \Bigl[\mbox{}^{\mu\nu} G^{\rho\sigma}\Bigr]\!(x;x') \,
\overline{\Psi}_{\!0}\!(x') \gamma_{\rho} J_{\sigma\beta}
\Psi_{\!0}\!(x') \times \gamma_{\mu} J_{\nu\alpha} \Psi_{\!0}\!(x)
\; .
\end{equation}
The term that contributes to the effective field equations is the
expectation value of the anti-commutator of $(4b)$ with
$b^{\dagger}(\vec{k},s)$,
\begin{eqnarray}
\lefteqn{\Bigl\langle \Omega \Bigl\vert
\Bigl\{(4b),b^{\dagger}\!(\vec{k},s) \Bigr\} \Bigr\vert \Omega
\Bigr\rangle = \frac{\kappa^2}4 \int d^Dx' \,
\partial^{\alpha} \partial^{\prime\beta} \Bigl[\mbox{}^{\mu\nu} G^{\rho\sigma}
\Bigr]\!(x;x') } \nonumber \\
& & \hspace{2.5cm} \times \Bigl\langle \Omega \Bigl\vert
\overline{\Psi}_{ \!0}\!(x') \gamma_{\rho} J_{\sigma\beta}
\Bigl\{\Psi_{\!0}\!(x'), b^{\dagger}\!(\vec{k},s)\Bigr\} \times
\gamma_{\mu} J_{\nu\alpha}
\Psi_{\!0}\!(x) \Bigr\vert \Omega \Bigr\rangle \; , \qquad \\
& & = \frac{\kappa^2}4 \int d^Dx' \, \partial^{\alpha}
\partial^{\prime\beta}
\Bigl[\mbox{}^{\mu\nu} G^{\rho\sigma}\Bigr]\!(x;x') \nonumber \\
& & \hspace{2.5cm} \times \Bigl\langle \Omega \Bigl\vert
\overline{\Psi}_{\!0}\!(x') \gamma_{\rho} J_{\sigma\beta} \,
\Xi_0\!(x') \times \gamma_{\mu} J_{\nu\alpha} \Psi_{\!0}\!(x)
\Bigr\vert \Omega \Bigr\rangle \; . \qquad
\end{eqnarray}
At this stage the spinor indices become confusing so we write them
out explicitly, and also remove all $\comp$-numbers from the
expectation value,
\begin{eqnarray}
\lefteqn{\Bigl\langle \Omega \Bigl\vert
\Bigl\{(4b)_i,b^{\dagger}\!(\vec{k},s) \Bigr\} \Bigr\vert \Omega
\Bigr\rangle = \frac{\kappa^2}4 \int d^Dx' \,
\partial^{\alpha} \partial^{\prime\beta} \Bigl[\mbox{}^{\mu\nu} G^{\rho\sigma}
\Bigr]\!(x;x') } \nonumber \\
& & \hspace{2.5cm} \times \Bigl\langle \Omega \Bigl\vert
\overline{\Psi}_{0k}(x') \Bigl(\gamma_{\rho}
J_{\sigma\beta}\Bigr)_{\!k\ell} \Xi_{0\ell}(x') \times
\Bigl(\gamma_{\mu} J_{\nu\alpha} \Bigr)_{\!ij}
\Psi_{\!0 j}\!(x) \Bigr\vert \Omega \Bigr\rangle \; , \qquad \\
& & = \frac{\kappa^2}4 \int d^Dx' \, \partial^{\alpha}
\partial^{\prime\beta}
\Bigl[\mbox{}^{\mu\nu} G^{\rho\sigma} \Bigr]\!(x;x') \times
\Bigl(\gamma_{\mu} J_{\nu\alpha} \Bigr)_{\!ij} \times
\Bigl(\gamma_{\rho}
J_{\sigma\beta}\Bigr)_{\!k\ell} \Xi_{0\ell}(x') \nonumber \\
& & \hspace{5 cm} \times \Bigl\langle \Omega \Bigl\vert\overline{
\Psi}_{\!0k}\!(x') \Psi_{\!0 j}\!(x) \Bigr\vert \Omega \Bigr\rangle
\; . \label{inter}
\end{eqnarray}

The expectation value on the final line of (\ref{inter}) is minus
the $+-$ fermion propagator,
\begin{equation}
\Bigl\langle \Omega \Bigl\vert\overline{\Psi}_{\!0k}\!(x')
\Psi_{\!0j}\!(x) \Bigr\vert \Omega \Bigr\rangle = - i\Bigl[\mbox{}_j
S_k\Bigr]_{\scriptscriptstyle \!+-}\!\!\!\!(x;x') = - i
\hspace{-.1cm}\not{\hspace{-.10cm} \mathcal{\partial}}_{jk} i
\Delta^{\rm cf}_{\scriptscriptstyle \!+-}\!\!(x \!-\! x') \; .
\label{prop}
\end{equation}
The minus sign derives from the fact that the preferred order for
the fermion propagator is $\Psi \overline{\Psi}$. Substituting
(\ref{prop}) into (\ref{inter}) gives an expression we can write
without resort to explicit spinor indices,
\begin{eqnarray}
\lefteqn{\Bigl\langle \Omega \Bigl\vert
\Bigl\{(4b),b^{\dagger}\!(\vec{k},s)
\Bigr\} \Bigr\vert \Omega \Bigr\rangle } \nonumber \\
& & \hspace{.4cm} = -\frac{i\kappa^2}4 \int d^Dx' \,
\partial^{\alpha}
\partial^{\prime\beta} \Bigl[\mbox{}^{\mu\nu} G^{\rho\sigma}\Bigr]\!(x;x')
\gamma_{\mu} J_{\nu\alpha} \hspace{-.1cm}\not{\hspace{-.10cm}
\mathcal{\partial}} i \Delta^{\rm cf}_{\scriptscriptstyle
\!+-}\!\!(x \!-\! x') \gamma_{\rho} J_{\sigma\beta} \Xi_{0}(x') \; .
\qquad \label{2nd}
\end{eqnarray}
Table~\ref{contribs} gives our results for each entry in
Table~\ref{idelPsi}.

\section{Our Rule}

The first eight entries of Table~\ref{contribs} provide a somewhat
cumbersome re-expres\-sion of the nonlocal contributions to the
order $\kappa^2$ source term of expression (\ref{source}). The
original source has the generic form of a difference of $++$ and
$+-$ terms, with each polarity being a product of contributions from
the graviton and contributions from the fermion.
Table~\ref{contribs} effects the following re-grouping,
\begin{eqnarray}
\lefteqn{\int d^Dx' \, \Bigl\{ ({\scriptscriptstyle ++})_{h} \times
({\scriptscriptstyle ++})_{\psi} - ({\scriptscriptstyle +-})_{h}
\times ({\scriptscriptstyle +-})_{\psi} \Bigr\} \, \Xi_0(x')
= \int d^Dx' \, ({\scriptscriptstyle ++})_{h} } \nonumber \\
& & \hspace{.5cm} \times \Bigl\{ ({\scriptscriptstyle ++})_{\psi} -
({\scriptscriptstyle +-})_{\psi} \Bigr\} \, \Xi_0(x') + \int d^Dx'
\Bigl\{({\scriptscriptstyle ++})_{h} - ({\scriptscriptstyle +-})_{h}
\Bigr\} \, ({\scriptscriptstyle +-})_{\psi} \Xi_0(x') \; . \qquad
\end{eqnarray}
From expression (\ref{Gret}) we see that the difference of $++$ and
$+-$ propagators for any field gives $-i$ times the retrarded
Green's function of that same field. The first eight entries come in
pairs of this form: $(1a)$-$(3a)$, $(1b)$-$(3b)$, $(2a)$-$(4a)$ and
$(2b)$-$(4b)$. This is an illuminating insight but it represents no
simplification of the original calculation.

We cannot simplify the propagators and retarded Green's functions
associated with the fermion. In contradistiction to the graviton,
the fermion is a ``passive'' field which cannot produce infrared
logarithms \cite{MW1,PTsW3}. Passive fields contribute factors of
order one that derive from both the infrared and the ultraviolet. To
correctly recover these factors the passive field must be treated
exactly.

Our simplification concerns the propagators and retarded Green's
functions of the graviton. The $A$-type graviton polarizations are
the ``active'' fields which cause infrared logarithms, whereas the
$B$-type and $C$-type polarizations are passive. Because this
particular calculation involves only one graviton propagator or
Green's function there is no chance of getting an infrared logarithm
unless the $A$-type part of the graviton propagator is involved.

Even within the $A$-type polarization, only the following tiny
portion of the infinite series expansion (\ref{DeltaA}) of
$i\Delta^{\!A}_{ \scriptscriptstyle \!+\pm}$ plays any role in
generating infrared logarithms,
\begin{eqnarray}
\lefteqn{i\delta\!\Delta^{\!A}_{\!+\pm}\!(x;x') \equiv
\frac{H^{D-2}}{(4 \pi)^{\frac{D}2}} \frac{\Gamma(D \!-\!
1)}{\Gamma(\frac{D}2)} \Biggl\{-\pi \cot\Bigl(\frac{D}2
\pi\Bigr) + \ln(a a')\Biggr\} } \nonumber \\
& & \hspace{6.5cm} + \frac{H^2}{8 \pi^{\frac{D}2}}
\frac{\Gamma(\frac{D}2 \!+\!1)}{D \!-\! 4} \frac{(a
a')^{2-\frac{D}2}}{\Delta x^{D-4}_{ \scriptscriptstyle \!+\pm}} \; .
\qquad\label{dAprop}
\end{eqnarray}
Our rule is accordingly to make the following simplifications on the
graviton propagators and Green's functions,
\begin{eqnarray}
i \Bigl[\mbox{}^{\mu\nu} \Delta^{\rho\sigma} \Bigr](x;x') &
\longrightarrow & \Bigl[\mbox{}^{\mu\nu} T_A^{\rho\sigma} \Bigr]
\times i\delta\!\Delta^{\!A}_{\scriptscriptstyle \!+\pm}\!(x;x') \;
,
\label{dAsimple} \\
\Bigl[\mbox{}^{\mu\nu} G^{\rho\sigma} \Bigr](x;x') & \longrightarrow
& \Bigl[\mbox{}^{\mu\nu} T_A^{\rho\sigma} \Bigr] \times \delta
\!G_{\!\!A}\!(x;x') \; , \label{GAsimple}
\end{eqnarray}
where the $A$-type tensor factor is (\ref{Atensor}) and we define
$\delta \!G_{\!\!A}\!(x;x')$ to be,
\begin{eqnarray}
\delta \!G_{\!\!A}\!(x;x') & \equiv & i \Bigl[
i\delta\!\Delta^{\!A}_{\scriptscriptstyle \!++}\!(x;x') -
i\delta\!\Delta^{\!A}_{\scriptscriptstyle \!+-}\!(x;x') \Bigr] \; , \\
& = & \frac{i H^2}{8 \pi^{\frac{D}2}} \frac{\Gamma(\frac{D}2
\!+\!1)}{ D \!-\! 4} \Bigl[\frac{(a a')^{2-\frac{D}2}}{\Delta
x^{D-4}_{\scriptscriptstyle \!++}} - \frac{(a
a')^{2-\frac{D}2}}{\Delta x^{D-4}_{\scriptscriptstyle \!+-}} \Bigr]
\; . \label{dG}
\end{eqnarray}
In the next section we demonstrate that applying our replacements
(\ref{dAsimple}-\ref{GAsimple}) to the various terms in
Table~\ref{contribs} reproduces the source term (\ref{keyeqn}) whose
integration gives the infrared logarithm (\ref{Z(t)}).

We close this section by commenting on the relation between our rule
and the replacements (\ref{srule1}-\ref{srule2}) that have been
shown to reproduce the leading infrared logarithms to all orders in
scalar models without derivative couplings \cite{MW1,SY,PTsW3}. The
rules are certainly not identical but they do seem to agree, at
leading logarithm order, for $D=4$ and for certain treatments of the
spatial coordinate separation. To see this, first take the $D=4$
limits of (\ref{dAprop}) and (\ref{dG}),
\begin{eqnarray}
\lim_{D \rightarrow 4} i\delta\!\Delta^{\!A}_{\!+\pm}\!(x;x') & = &
-\frac{H^2}{8 \pi^2} \Biggl\{ \ln\Bigl[\frac14 H^2 \Delta
x^2_{\scriptscriptstyle \!+\pm} \Bigr] + \frac12 \Biggr\} \; , \label{DA4} \\
\lim_{D \rightarrow 4} \delta \!G_{\!\!A}\!(x;x') & = & \frac{H^2}{4
\pi} \, \theta\Bigl(\eta \!-\! \eta' \!-\! \Vert \vec{x} \!-\!
\vec{x}'\Vert\Bigr) \; . \label{DG4}
\end{eqnarray}
If we set $\vec{x}' = \vec{x}$ in (\ref{DA4}) the result is,
\begin{equation}
-\frac{H^2}{8 \pi^2} \Biggl\{ \ln\Bigl[\frac14 \Bigl(\frac1{a'}
\!-\! \frac1{a}\Bigr)^2\Bigr] + \frac12\Biggr\} = \frac{H^2}{4
\pi^2} \Biggl\{ \ln\Bigl[ {\rm min}(a,a')\Bigr] + O(1)\Biggr\} \; .
\end{equation}
At leading logarithm order this indeed agrees with the $D=4$ limit
of our previous rule (\ref{srule1}). Similarly, the spatial integral
of (\ref{DG4}) is,
\begin{equation}
\int d^3 x \, \frac{H^2}{4\pi} \, \theta\Bigl(\eta \!-\! \eta' \!-\!
\Vert \vec{x} \!-\! \vec{x}'\Vert\Bigr) = \frac1{3 H}
\Bigl(\frac1{a'} \!-\! \frac1{a}\Bigr)^3 = \frac1{3 H a^{\prime 3}}
\, \Biggl[1 + O\Bigl(\frac{a'}{a}\Bigr) \Biggr] \; .
\end{equation}
At leading logarithm order this agrees with the $D=4$ limit of the
spatial integral of (\ref{srule2}).

These correspondences seem to mean that our new replacements
(\ref{dAsimple}-\ref{GAsimple}) would reproduce the leading infrared
logarithms of the simple scalar models previously studied. However,
it is straightforward to check that the old replacements
(\ref{srule1}-\ref{srule2}) do {\it not} reproduce the result
(\ref{keyeqn}) we get from quantum gravity, whereas our new
replacements (\ref{dAsimple}-\ref{GAsimple}) do. It therefore seems
that our new rule represents a successful generalization of the old
rule to the more singular environment that arises when derivative
couplings are present. What is not yet clear is whether or not the
rule can be simplified.

\section{Analysis}

The purpose of this section is to show that applying our rule
(\ref{dAsimple}-\ref{GAsimple}) to Table~\ref{contribs} reproduces
the result (\ref{keyeqn}) of our explicit computation. We begin by
observing that any terms involving derivatives of $\Xi_0$ cannot
contribute at leading order. That reduces the problem to considering
the nonlocal contributions $1b$, $2b$, $3b$ and $4b$, and the local
contributions $6$, $7$ and $8$. The local contributions were
evaluated in an earlier effort to understand our result
(\ref{keyeqn}) on a qualitative level by making the Hartree
approximation \cite{MW3}, so we concentrate on the nonlocal
contributions. We first introduce a systematic classification for
the myriads of distinct terms they give when the $A$-type tensor
factor and the factors of $\gamma_{\mu} J_{\nu\alpha}$ and
$\gamma_{\rho} J_{\sigma\beta}$ are broken up. Then we explicitly
evaluate four of the contributions from $(2b)$ as an example. Final
results for all nonlocal and local contributions are reported in
tables.

It is important to understand that we only seek the leading late
time behaviors of the various source terms in Table~\ref{contribs}.
By considering the form of quantum gravity interactions we see that
the one loop mode function can be enhanced by at most a single
infrared logarithm \cite{PTsW3},
\begin{equation}
\kappa^2 \Bigl\langle \Omega \Bigl\vert \Bigl\{ \Psi_2(x),
b^{\dagger}(\vec{k},s) \Bigr\} \Bigr\vert \Omega \Bigr\rangle \sim
\kappa^2 H^2 \times \ln(a) \times \Xi_0(x) \; .
\end{equation}
The source terms of Table~\ref{contribs} should be $i
\hspace{-.15cm} \not{\hspace{-.10cm} \partial}$ times this, which
gives the loop counting parameter $\kappa^2 H^2$ times $i a H
\gamma^0 \Xi_0(x)$.

Now consider how the derivatives of Table~\ref{contribs} act. Any
which act on the tree order mode function $\Xi_0(x')$ will bring
down factors of the wave number $k$. This factor of $k$ must
persist, even after the integration over $x^{\prime \mu}$, because
the integral remains finite for $\vec{k} = 0$. Further, this factor
of $k$ will always be accompanied by a factor of $1/a$ to make the
wave number physical. It follows that the fastest growth possible
for any $\partial' \Xi_0(x')$ term is $\ln(a) k \gamma^0 \Xi_0(x)$.
We can therefore forget about nonlocal contributions from $(1a)$,
$(2a)$, $(3a)$ or $(4a)$, and also the local contribution from
$(5)$. For the same reason we can make the following simplification
in the nonlocal contributions from $(1b)$, $(2b)$, $(3b)$ and
$(4b)$,
\begin{equation}
\Xi_0(x';\vec{k},s) = \Xi_0(x;\vec{k},s) \times e^{-i k_{\mu} (x -
x')^{\mu}} \longrightarrow \Xi_0(x;\vec{k},s) \times 1 \; .
\end{equation}

We turn now to the problem of classifying the many distinct
contributions that derive from $(1b)$, $(2b)$, $(3b)$ and $(4b)$.
These four terms all involve a single factor of the $A$-type tensor
and either one or two factors of the Lorentz generators. The
$A$-type tensor indices are purely spatial, for example,
\begin{equation}
\Bigl[\mbox{}^{\mu\nu} T_A^{\rho\sigma}\Bigr] \times
\Bigl(\gamma_{\mu} J_{\nu\alpha}\Bigr) \times \Bigl(\gamma_{\rho}
J_{\sigma \beta}\Bigr) = \Bigl[\mbox{}^{ij} T_A^{k \ell}\Bigr]
\times \Bigl(\gamma_i J_{j\alpha}\Bigr) \times \Bigl(\gamma_k
J_{\ell \beta}\Bigr) \; . \label{strucs}
\end{equation}
Our classification system is based upon decomposing the $\gamma
\cdot J$ factors as follows,
\begin{eqnarray}
\gamma_i J_{j\beta} = \frac{i}2 \gamma_{[i} \gamma_j
\gamma_{\alpha]} + \frac{i}2 \delta_{i\alpha} \gamma_j - \frac{i}2
\delta_{ij} \gamma_{\alpha} \; .
\end{eqnarray}
The totally anti-symmetrized term drops out because the $A$-type
tensor factor is symmetric in $i$ and $j$. We label the other two
terms by Roman numerals ``I'' and ``II'' as follows,
\begin{eqnarray}
{\rm I} & \Longleftrightarrow & \frac{i}2 \delta_{i\alpha} \gamma_j
\qquad {\rm and} \qquad \frac{i}2 \delta_{k\beta} \gamma_{\ell} \; , \\
{\rm II} & \Longleftrightarrow & \frac{i}2 \delta_{ij}
\gamma_{\alpha} \qquad {\rm and} \qquad \frac{i}2 \delta_{k\ell}
\gamma_{\beta} \; .
\end{eqnarray}
The indices $\alpha$ --- which appears only in $(2b)$ and $(4b)$ ---
and $\beta$, contract into derivative operators $\partial^{\alpha}$
and $\partial^{\prime \beta}$ that act upon the graviton propagator
or retarded Green's function. The indices on I-type terms must be
spatial but those on II-type terms can be either spatial --- denoted
by just II --- or temporal --- denoted by ${\rm II}'$. The type-II
term always produces a contraction of the $A$-type tensor factor,
for example,
\begin{equation}
\Bigl[\mbox{}^{ij} T_A^{k\ell}\Bigr] \times \delta_{k\ell} =
-\frac4{D\!-\!3} \, \delta^{ij} \; .
\end{equation}
However, the type-I term can receive distinct contributions from
each of the three terms in $[\mbox{}^{ij} T_A^{k\ell}]$. Where the
results are distinct we label these ``A'', ``B'' and ``C'' as
follows,
\begin{equation}
{\rm A} \Longleftrightarrow \delta^{ik} \delta^{j\ell} \qquad ,
\qquad {\rm B} \Longleftrightarrow \delta^{i\ell} \delta^{j k}
\qquad , \qquad {\rm C} \Longleftrightarrow -\frac2{D\!-\!3}
\delta^{i j} \delta^{k\ell} \; .
\end{equation}

The various classifications are arranged in a prescribed order, and
are separated by periods. First comes the term designation --- 1b,
2b, 3b or 4b. Next comes the leftmost of the $\gamma \cdot J$ factor
designations
--- I, II or ${\rm II}'$. If there is a second $\gamma \cdot J$ factor,
its designator comes next. The final designator is the A, B, or C
from the tensor factor, with no designator denoting the presence of
all three terms. As an example, consider the full $(1b)$ term,
\begin{equation}
\frac{\kappa^2}4 \int \! d^Dx' \, \partial^{\prime \beta} i \delta\!
\Delta^A_{\scriptscriptstyle \!++}\!(x;x') \, \Bigl[\mbox{}^{ij}
T_A^{k\ell} \Bigr] \gamma_i \partial_j \hspace{-.1cm}
\not{\hspace{-.1cm}
\partial} \,
G_{\rm cf}(x\!-\!x') \gamma_k J_{\ell\beta} \Xi_0(x') \; .
\end{equation}
The ${\rm 1b.II'}$ contribution is,
\begin{eqnarray}
{\rm 1b.II'} & = & \frac{\kappa^2}4 \int \! d^Dx' \times
\Bigl(\partial^{\prime 0}\Bigr) \times i \delta\!
\Delta^A_{\scriptscriptstyle \!++}\!(x;x') \times \Bigl(
-\frac4{D\!-\!3} \, \delta^{ij} \Bigr) \nonumber \\
& & \hspace{3cm} \times \gamma_i \partial_j \hspace{-.1cm}
\not{\hspace{-.1cm} \partial} \, G_{\rm cf}(x\!-\!x') \times
\Bigl(- \frac{i}2 \gamma_0 \Bigr) \times \Xi_0(x') \; , \qquad \\
& = & \frac{i \kappa^2}{2 (D-3)} \!\int\!d^{D}x' \, \partial'_0
i\delta \! \Delta^A_{\scriptscriptstyle \!++}\!(x;x') \hspace{-.1cm}
\not{\hspace{-.1cm} \overline{\partial}} \hspace{-.1cm}
\not{\hspace{-.1cm} \partial} \, G_{\rm cf}(\!x\!-\!x') \gamma^0
\Xi_0(x') \; .
\end{eqnarray}
In contrast, the contribution from 1b.I.B is,
\begin{eqnarray}
{\rm 1b.I.B} & = & \frac{\kappa^2}4 \int \! d^Dx' \times
\Bigl(\partial_k' \Bigr) \times i \delta\!
\Delta^A_{\scriptscriptstyle \!++}\!(x;x') \times \Bigl(
\delta^{i\ell} \delta^{jk} \Bigr) \nonumber \\
& & \hspace{3cm} \times \gamma_i \partial_j \hspace{-.1cm}
\not{\hspace{-.1cm}
\partial} \, G_{\rm cf}(x\!-\!x') \times \Bigl(\frac{i}2 \gamma_{\ell} \Bigr)
\times \Xi_0(x') \; , \qquad \\
& & = -\frac{i \kappa^2}8 \!\int\!d^Dx' \, \partial_k i\delta
\!\Delta^A_{ \scriptscriptstyle \!++}\!(x;x') \gamma^{\ell}
\partial_k \hspace{-.1cm}
\not{\hspace{-.1cm}\partial}\, G_{\rm cf}(\!x\!-\!x') \gamma^{\ell}
\Xi_0(x') \; .
\end{eqnarray}
Note the minus sign from converting $\partial_k'$ to $-\partial_k$.

\begin{table}

\vbox{\tabskip=0pt \offinterlineskip
\def\tablerule{\noalign{\hrule}}
\halign to390pt {\strut#& \vrule#\tabskip=1em plus2em& \hfil#\hfil&
\vrule#& \hfil#\hfil& \vrule#& \hfil#\hfil& \vrule#\tabskip=0pt\cr
\tablerule \omit&height4pt&\omit&&\omit&&\omit&\cr
&&$\!\!\!\!{\rm\,Term}\!\!\!\!$ && ${\rm Contribution\,\, from\,\,}
\Bigl\langle \Omega \Bigl\vert
\Bigl\{(1b),b^{\dagger}(\vec{k},s)\Bigr\} \Bigr\vert \Omega
\Bigr\rangle$ && $\!\!\!\!\!{\rm Coef.}\!\!\!\!\!$ & \cr
\omit&height4pt&\omit&&\omit&&\omit&\cr \tablerule
\omit&height4pt&\omit&&\omit&&\omit&\cr && $\!\!\!\!{\rm
1b.II'}\!\!\!\!$ && $\frac{i \kappa^2}{2 (D-3)} \!\int\!d^{D}x'
\partial'_0 i\delta \!\Delta^A_{\!++}\!(x;x')
\hspace{-.1cm}\not{\hspace{-.1cm}\overline{\partial}}
\hspace{-.1cm}\not{\hspace{-.1cm}\partial}\,G_{\rm cf}(\!x\!-\!x')
\gamma^{0} \Xi_{0}(x') $ && $\!\!\!\!\!+\frac14 \!\!\!\!\!$ & \cr
\omit&height4pt&\omit&&\omit&&\omit&\cr \tablerule
\omit&height4pt&\omit&&\omit&&\omit&\cr &&$\!\!\!\!{\rm
1b.II}\!\!\!\!$ && $-\frac{i \kappa^2}{2 (D-3)} \!\int\!d^{D}x'
\partial_k i\delta \!\Delta^A_{\!++}\!(x;x')
\hspace{-.1cm}\not{\hspace{-.1cm}\overline{\partial}} \hspace{-.1cm}
\not{\hspace{-.1cm}\partial}\,G_{\rm cf}(\!x\!-\!x') \gamma^k
\Xi_{0}(x')$ && $\!\!\!\!\!-\frac14 \!\!\!\!\!$ & \cr
\omit&height4pt&\omit&&\omit&&\omit&\cr \tablerule
\omit&height4pt&\omit&&\omit&&\omit&\cr && $\!\!\!\!{\rm
1b.I.A}\!\!\!\!$ && $-\frac{i \kappa^2}8 \!\int\!d^Dx'
\partial_k i\delta\!\Delta^A_{\!++}\!(x;x')\gamma^k \partial_{\ell}
\hspace{-.1cm} \not{\hspace{-.1cm}\partial} \,G_{\rm cf}(\!x\!-\!x')
\gamma^{\ell} \Xi_0(x')$ && $\!\!\!\!\!-\frac1{16}\!\!\!\!\!$ & \cr
\omit&height4pt&\omit&&\omit&&\omit&\cr \tablerule
\omit&height4pt&\omit&&\omit&&\omit&\cr && $\!\!\!\!{\rm
1b.I.B}\!\!\!\!$ && $-\frac{i \kappa^2}8 \!\int\!d^Dx'
\partial_k i\delta \!\Delta^A_{\!++}\!(x;x') \gamma^{\ell} \partial_k
\hspace{-.1cm} \not{\hspace{-.1cm}\partial}\, G_{\rm cf}(\!x\!-\!x')
\gamma^{\ell} \Xi_0(x')$ && $\!\!\!\!\!-\frac3{16}\!\!\!\!\!$ & \cr
\omit&height4pt&\omit&&\omit&&\omit&\cr \tablerule
\omit&height4pt&\omit&&\omit&&\omit&\cr &&$\!\!\!\!{\rm
1b.I.C}\!\!\!\!$ && $\frac{i \kappa^2}{4 (D-3)}\!\int\!d^Dx'
\partial_k i\delta \!\Delta^A_{\!++}\!(x;x')
\hspace{-.1cm}\not{\hspace{-.1cm}\overline{\partial}} \hspace{-.1cm}
\not{\hspace{-.1cm}\partial} \, G_{\rm cf}(\!x\!-\!x') \gamma^k
\Xi_{0}(x')$ && $\!\!\!\!\!+\frac18 \!\!\!\!\!$ & \cr
\omit&height4pt&\omit&&\omit&&\omit&\cr \tablerule
\omit&height2pt&\omit&&\omit&&\omit&\cr \tablerule
\omit&height4pt&\omit&&\omit&&\omit&\cr &&$\!\!\!\!{\rm
Total}\!\!\!\!$ && $\!\!\!\!\!\frac{\kappa^2}4 \!\int\!d^Dx'
\partial^{\prime \beta} i\delta\! \Delta^A_{\!++}\!(x;x') [\mbox{}^{ij}
T_A^{k\ell}] \gamma_i \partial_j \! \hspace{-.1cm}
\not{\hspace{-.1cm}\partial} G_{\rm cf}(\!x\!-\!x') \gamma_k J_{\ell
\beta} \Xi_{0}(x')\!\!\!\!\!$ && $\!\!\!\!\!-\frac18 \!\!\!\!\!$ &
\cr \omit&height4pt&\omit&&\omit&&\omit&\cr \tablerule}}

\caption{The full contribution from each $(1b)$ term consists of its
numerical coefficient times $\frac{i\kappa^2H^2}{16
\pi^2}Ha\gamma^{0}\Xi_{0}(x)$.} \label{table1b}
\end{table}

To describe the evaluation technique we have chosen four of the
contributions from $(2b)$: ${\rm 2b.II'.I}$, 2b.II.I, ${\rm
2b.I.II'}$ and 2b.I.II. Each of these involves a single derivative
of the conformal Green's function,
\begin{equation}
\hspace{-.1cm} \not{\hspace{-.1cm}\partial} G_{\rm cf}(x \!-\! x') =
\frac{i \Gamma(\frac{D}2)}{2 \pi^{\frac{D}2}} \,\Bigl[-\frac1{\Delta
x^D_{ \scriptscriptstyle \!++}} + \frac1{\Delta
x^D_{\scriptscriptstyle \!+-}}\Bigr] \gamma^{\mu} \Delta x_{\mu} \;
.
\end{equation}
They also involve two derivatives of the $A$-type propagator,
\begin{eqnarray}
\partial_k \partial_0' i\delta \!\Delta^{\!A}_{\scriptscriptstyle \!+\pm}\!
(x;x') & \!\!=\!\! & \frac{H^2 \Gamma(\frac{D}2 \!+\!1)}{8
\pi^{\frac{D}{2}} (a a')^{\frac{D}2 - 2}} \, \Delta x^k \Biggl\{
\frac{(D\!-\!2) \Delta \eta}{ \Delta x^D_{\scriptscriptstyle
\!+\pm}} + \frac{(D\!-\!4) H a'}{2 \Delta
x^{D-2}_{\scriptscriptstyle \!+\pm}} \Biggr\} \; , \qquad \\
\partial_k \partial_0 i\delta \!\Delta^{\!A}_{\scriptscriptstyle \!+\pm}\!
(x;x') & \!\!=\!\! & \frac{H^2 \Gamma(\frac{D}2 \!+\!1)}{8
\pi^{\frac{D}{2}} (a a')^{\frac{D}2 - 2}} \, \Delta x^k \Biggl\{-
\frac{(D\!-\!2) \Delta \eta}{ \Delta x^D_{\scriptscriptstyle
\!+\pm}} + \frac{(D\!-\!4) H a}{2 \Delta
x^{D-2}_{\scriptscriptstyle \!+\pm}} \Biggr\} \; , \qquad \\
\partial_k \partial_{\ell} i\delta \!\Delta^{\!A}_{\scriptscriptstyle \!+\pm}\!
(x;x') & \!\!=\!\! & \frac{H^2 \Gamma(\frac{D}2 \!+\!1)}{8
\pi^{\frac{D}{2}} (a a')^{\frac{D}2 - 2}} \, \Biggl\{
-\frac{\delta^{k\ell}}{\Delta x^{D-2}_{ \scriptscriptstyle \!+\pm}}
+ \frac{(D\!-\!2) \Delta x^k \Delta x^{\ell}}{ \Delta
x^D_{\scriptscriptstyle \!+\pm}} \Biggr\} \; .
\end{eqnarray}
In these and all subsequent expressions we define the coordinate
differences,
\begin{equation}
\Delta x^{\mu} \equiv x^{\mu} - x^{\prime \mu} \qquad {\rm and}
\qquad \Delta \eta \equiv \eta - \eta' \; .
\end{equation}

Each of the four terms we are considering takes the form of a common
integral operator acting upon a different integrand. The integral
operator is,
\begin{equation}
\frac{\kappa^2 H^2}{64 \pi^{D}} \frac{\Gamma(\frac{D}2)
\Gamma(\frac{D}2 \!+\! 1)}{D - 3} \int d^Dx' \, (a a')^{2 -
\frac{D}2} \Biggl[-\frac1{\Delta x^D_{\scriptscriptstyle ++}}
+\frac1{\Delta x^D_{\scriptscriptstyle +-}} \Biggr] \, \times \; .
\end{equation}
The four different integrands are,
\begin{eqnarray}
{\rm 2b.II'.I} & \Longrightarrow & \gamma^0 \gamma^{\mu} \gamma^k
\Xi_0(x') \times \Delta x_{\mu} \Delta x^k \Biggl\{- \frac{(D\!-\!2)
\Delta \eta}{\Delta x^D_{\scriptscriptstyle ++}} + \frac{(D\!-\!4) H
a}{2 \Delta x^{D-2}_{
\scriptscriptstyle ++}} \Biggr\} \; , \qquad \\
{\rm 2b.II.I} & \Longrightarrow & \gamma^k \gamma^{\mu}
\gamma^{\ell} \Xi_0(x') \times \Delta x_{\mu}
\Biggl\{-\frac{\delta^{k\ell}}{\Delta x^{D-2}_{ \scriptscriptstyle
++}} + \frac{(D\!-\!2) \Delta x^k \Delta x^{\ell}}{\Delta
x^{D}_{\scriptscriptstyle ++}} \Biggr\} \; , \\
{\rm 2b.I.II'} & \Longrightarrow & \gamma^k \gamma^{\mu} \gamma^0
\Xi_0(x') \times \Delta x_{\mu} \Delta x^k \Biggl\{- \frac{(D\!-\!2)
\Delta \eta}{\Delta x^D_{\scriptscriptstyle ++}} - \frac{(D\!-\!4) H
a'}{2 \Delta x^{D-2}_{
\scriptscriptstyle ++}} \Biggr\} \; , \qquad \\
{\rm 2b.I.II} & \Longrightarrow & \gamma^k \gamma^{\mu}
\gamma^{\ell} \Xi_0(x') \times \Delta x_{\mu}
\Biggl\{-\frac{\delta^{k\ell}}{\Delta x^{D-2}_{ \scriptscriptstyle
++}} + \frac{(D\!-\!2) \Delta x^k \Delta x^{\ell}}{\Delta
x^{D}_{\scriptscriptstyle ++}} \Biggr\} \; .
\end{eqnarray}
If we ignore the difference between $x^{\prime \mu}$ and $x^{\mu}$
in the wavefunction $\Xi_0(x')$ and perform the angular averages,
the various integrands take the form,
\begin{eqnarray}
{\rm 2b.II'.I} & \Longrightarrow & \gamma^0 \Xi_0(x) \times
\Biggl\{\frac{(D \!-\!2) \Delta \eta \Vert \Delta
\vec{x}\Vert^2}{\Delta x^{D}_{ \scriptscriptstyle ++}} -
\frac{(D\!-\!4) H a \Vert \Delta \vec{x}\Vert^2}{2
\Delta x^{D-2}_{\scriptscriptstyle ++}} \Biggr\} \; , \\
{\rm 2b.II.I} & \Longrightarrow & \gamma^0 \Xi_0(x) \times \Biggl\{
\frac{(D\!-\!1) \Delta \eta}{\Delta x^{D-2}_{\scriptscriptstyle ++}}
- \frac{(D\!-\!2) \Delta \eta \Vert \Delta \vec{x}\Vert^2}{\Delta
x^{D}_{
\scriptscriptstyle ++}} \Biggr\} \; , \\
{\rm 2b.I.II'} & \Longrightarrow & \gamma^0 \Xi_0(x) \times \Biggl\{
\frac{(D\!-\!2) \Delta \eta \Vert \Delta \vec{x}\Vert^2}{\Delta
x^{D}_{ \scriptscriptstyle ++}} + \frac{(D\!-\!4) H a' \Vert \Delta
\vec{x}\Vert^2}{2
\Delta x^{D-2}_{\scriptscriptstyle ++}} \Biggr\} \; , \qquad \\
{\rm 2b.I,II} & \Longrightarrow & \gamma^0 \Xi_0(x) \times \Biggl\{
\frac{(D\!-\!1) \Delta \eta}{\Delta x^{D-2}_{\scriptscriptstyle ++}}
- \frac{(D\!-\!2) \Delta \eta \Vert \Delta \vec{x}\Vert^2}{\Delta
x^{D}_{ \scriptscriptstyle ++}} \Biggr\} \; .
\end{eqnarray}

\begin{table}

\vbox{\tabskip=0pt \offinterlineskip
\def\tablerule{\noalign{\hrule}}
\halign to400pt {\strut#& \vrule#\tabskip=1em plus2em& \hfil#\hfil&
\vrule#& \hfil#\hfil& \vrule#& \hfil#\hfil& \vrule#\tabskip=0pt\cr
\tablerule \omit&height4pt&\omit&&\omit&&\omit&\cr
&&$\!\!\!\!{\rm\,Term}\!\!\!\!$ && ${\rm Contribution\,\, from\,\,}
\Bigl\langle \Omega \Bigl\vert
\Bigl\{(3b),b^{\dagger}(\vec{k},s)\Bigr\} \Bigr\vert \Omega
\Bigr\rangle$ && $\!\!\!\!\!{\rm Coef.}\!\!\!\!\!$ & \cr
\omit&height4pt&\omit&&\omit&&\omit&\cr \tablerule
\omit&height4pt&\omit&&\omit&&\omit&\cr &&$\!\!\!\!{\rm
3b.II'}\!\!\!\!$ && $\frac{i \kappa^2}{2 (D-3)} \int\!d^Dx'
\partial'_0 \delta \!G_{\!\!A}\!(x;x')
\hspace{-.1cm}\not{\hspace{-.1cm}\overline{\partial}} \hspace{-.1cm}
\not{\hspace{-.1cm}{\partial}} \, i\Delta^{\rm cf}_{+-}\!(x\!-\!x')
\gamma^0 \Xi_{0}(x')$ && $+\frac14$ & \cr
\omit&height4pt&\omit&&\omit&&\omit&\cr \tablerule
\omit&height4pt&\omit&&\omit&&\omit&\cr &&$\!\!\!\!{\rm
3b.II}\!\!\!\!$ && $-\frac{i \kappa^2}{2 (D-3)} \int\!d^Dx'
\partial_k \delta \!G_{\!\!A}\!(x;x') \hspace{-.1cm} \not{\hspace{-.1cm}
\overline{\partial}} \hspace{-.1cm}\not{\hspace{-.1cm}{\partial}} \,
i\Delta^{\rm cf}_{+-}\!(x\!-\!x') \gamma^k \Xi_{0}(x')$ &&
$-\frac14$ & \cr \omit&height4pt&\omit&&\omit&&\omit&\cr \tablerule
\omit&height4pt&\omit&&\omit&&\omit&\cr &&$\!\!\!\!{\rm
3b.I.A}\!\!\!\!$ && $-\frac{i \kappa^2}8 \int\!d^Dx'
\partial_k
\delta \!G_{\!\!A}\!(x;x') \gamma^k \partial_{\ell} \hspace{-.1cm}
\not{\hspace{-.1cm}{\partial}} \, i\Delta^{\rm cf}_{+-}\!(x\!-\!x')
\gamma^{\ell} \Xi_{0}(x')$ && $-\frac1{16}$ & \cr
\omit&height4pt&\omit&&\omit&&\omit&\cr \tablerule
\omit&height4pt&\omit&&\omit&&\omit&\cr &&$\!\!\!\!{\rm
3b.I.B}\!\!\!\!$ && $-\frac{i \kappa^2}8 \int\!d^Dx'
\partial_k
\delta \!G_{\!\!A}\!(x;x') \gamma^{\ell} \partial_k \hspace{-.1cm}
\not{\hspace{-.1cm}{\partial}} \, i\Delta^{\rm cf}_{+-}\!(x\!-\!x')
\gamma^{\ell} \Xi_{0}(x')$ && $-\frac3{16}$ & \cr
\omit&height4pt&\omit&&\omit&&\omit&\cr \tablerule
\omit&height4pt&\omit&&\omit&&\omit&\cr &&$\!\!\!\!{\rm
3b.I.C}\!\!\!\!$ && $\frac{i \kappa^2}{4 (D-3)} \int\!d^Dx'
\partial_k \delta \!G_{\!\!A}\!(x;x')
\hspace{-.1cm}\not{\hspace{-.1cm}\overline{\partial}} \hspace{-.1cm}
\not{\hspace{-.1cm}{\partial}} \, i\Delta^{\rm cf}_{+-}\!(x\!-\!x')
\gamma^k \Xi_{0}(x')$ && $+\frac18$ & \cr
\omit&height4pt&\omit&&\omit&&\omit&\cr \tablerule
\omit&height2pt&\omit&&\omit&&\omit&\cr \tablerule
\omit&height4pt&\omit&&\omit&&\omit&\cr &&$\!\!\!\!{\rm
Total}\!\!\!\!$ && $\!\!\!\!\!\frac{\kappa^2}4 \!\int\!d^Dx'
\partial^{\prime \beta} \delta \! G_{\!\!A}\!(x;x') [\mbox{}^{ij} T_A^{k\ell}]
\gamma_i \partial_j \! \hspace{-.1cm} \not{\hspace{-.1cm}\partial}
i\Delta^{\rm cf}_{+-}\!(\!x\!-\!x') \gamma_k J_{\ell \beta}
\Xi_{0}(x') \!\!\!\!\!$ && $\!\!\!\!\!-\frac18 \!\!\!\!\!$ & \cr
\omit&height4pt&\omit&&\omit&&\omit&\cr \tablerule}}

\caption{The full contribution from each $(3b)$ term consists of its
numerical coefficient times $\frac{i \kappa^2 H^2}{16 \pi^2} H a
\gamma^0 \Xi_0(x)$.} \label{table3b}
\end{table}

Each of these four terms can be written as a common factor times a
sum of integrals. The common factor is,
\begin{equation}
\frac{\kappa^2 H^2}{64 \pi^{\frac{D}2+2}} \frac{\Gamma(\frac{D}2)
\Gamma(\frac{D}2 \!+\! 1)}{D - 3} \, \gamma^0 \Xi_0(x) \; .
\end{equation}
The four fundamental integrals are,
\begin{eqnarray}
I_1 & \equiv & (D\!-\!2) \int d^Dx' \, (a a')^{2 - \frac{D}2}
\Biggl[- \frac1{\Delta x^D_{\scriptscriptstyle ++}} +\frac1{\Delta
x^D_{ \scriptscriptstyle +-}}\Biggr] \frac{\Delta \eta \Vert \Delta
\vec{x}
\Vert^2}{\Delta x^D_{\scriptscriptstyle ++}} \; , \\
I_2 & \equiv & \frac12 (D\!-\!4) \int d^Dx' \, (a a')^{2 -
\frac{D}2} \Biggl[- \frac1{\Delta x^D_{\scriptscriptstyle ++}}
+\frac1{\Delta x^D_{ \scriptscriptstyle +-}}\Biggr] \frac{H a \Vert
\Delta \vec{x} \Vert^2}{
\Delta x^{D-2}_{\scriptscriptstyle ++}} \; , \\
I_3 & \equiv & (D\!-\!1) \int d^Dx' \, (a a')^{2 - \frac{D}2}
\Biggl[- \frac1{\Delta x^D_{\scriptscriptstyle ++}} +\frac1{\Delta
x^D_{ \scriptscriptstyle +-}}\Biggr] \frac{\Delta \eta}{\Delta
x^{D-2}_{
\scriptscriptstyle ++}} \; , \\
I_4 & \equiv & \frac12 (D\!-\!4) \int d^Dx' \, (a a')^{2 -
\frac{D}2} \Biggl[- \frac1{\Delta x^D_{\scriptscriptstyle ++}}
+\frac1{\Delta x^D_{ \scriptscriptstyle +-}}\Biggr] \frac{H a' \Vert
\Delta \vec{x} \Vert^2}{ \Delta x^{D-2}_{\scriptscriptstyle ++}} \;
.
\end{eqnarray}
And the four terms under consideration are,
\begin{eqnarray}
{\rm 2b.II'.I} & \Longrightarrow & \frac{\kappa^2 H^2}{64 \pi^{D}}
\frac{\Gamma(\frac{D}2) \Gamma(\frac{D}2 \!+\! 1)}{D - 3} \,
\gamma^0 \Xi_0(x)
\times \Bigl\{I_1 - I_2\Bigr\} \; , \label{0,n} \\
{\rm 2b.II.I} & \Longrightarrow & \frac{\kappa^2 H^2}{64 \pi^{D}}
\frac{\Gamma(\frac{D}2) \Gamma(\frac{D}2 \!+\! 1)}{D - 3} \,
\gamma^0 \Xi_0(x)
\times \Bigl\{I_3 - I_1\Bigr\} \; , \label{m,n} \\
{\rm 2b.I.II'} & \Longrightarrow & \frac{\kappa^2 H^2}{64 \pi^{D}}
\frac{\Gamma(\frac{D}2) \Gamma(\frac{D}2 \!+\! 1)}{D - 3} \,
\gamma^0 \Xi_0(x)
\times \Bigl\{I_1 + I_4\Bigr\} \; , \label{m,0} \\
{\rm 2b.I.II} & \Longrightarrow & \frac{\kappa^2 H^2}{64 \pi^{D}}
\frac{\Gamma(\frac{D}2) \Gamma(\frac{D}2 \!+\! 1)}{D - 3} \,
\gamma^0 \Xi_0(x) \times \Bigl\{I_3 - I_1\Bigr\} \; . \label{mn}
\end{eqnarray}
The procedure for evaluating $I_{1-4}$ is,
\begin{enumerate}
\item{Perform the angular integrations;}
\item{Note that (for $\delta \rightarrow 0$) the radial integrand
vanishes for $r > \Delta \eta$;}
\item{Make the change of variable $r = \Delta \eta \sqrt{x}$, which
reduces the radial integrals to beta functions; and}
\item{Make the change of variable $\eta' = -1/(H a t)$.}
\end{enumerate}
As an example, consider $I_1$. The first step brings it to,
\begin{eqnarray}
\lefteqn{I_1 = (D\!-\!2) \times \frac{2
\pi^{\frac{D-1}2}}{\Gamma(\frac{D-1}2)} \times
\int_{-\frac1{H}}^{\eta} \!\!\! d\eta' \, (a a')^{2-\frac{D}2}
\int_0^{\infty} \!\!\! dr \, r^{D-2} } \nonumber \\
& & \times \Biggl\{-\frac1{[r^2 - (\Delta \eta \!-\! i
\delta)^2]^{\frac{D}2}} + \frac1{[r^2 - (\Delta \eta \!+\! i
\delta)^2]^{\frac{D}2}} \Biggr\} \frac{\Delta \eta \, r^2}{[r^2 -
(\Delta \eta \!-\! i \delta)^2]^{\frac{D}2}} \; . \qquad
\label{step1}
\end{eqnarray}
Step 2 is accomplished by noting that the $\mp i \delta$ factors
serve to fix the phase of the complex numbers that must be raised to
the $D/2$ power on the final line of (\ref{step1}). For $r > \Delta
\eta$ that phase is zero, whereas it is $\pm \pi$ for $0 < r <
\Delta \eta$,
\begin{equation}
r^2 - (\Delta \eta \mp i \delta)^2 = e^{\pm i \pi} \times
\Bigl(\Delta \eta^2 - r^2\Bigr) \qquad {\rm for} \qquad 0 < r <
\Delta \eta \; .
\end{equation}
Hence the curly bracketed term of (\ref{step1}) becomes,
\begin{equation}
-\frac1{[r^2 - (\Delta \eta \!-\! i \delta)^2]^{\frac{D}2}} +
\frac1{[r^2 - (\Delta \eta \!+\! i \delta)^2]^{\frac{D}2}} = \frac{2
i \sin(\frac{\pi D}2)}{[\Delta \eta^2 - r^2]^{\frac{D}2}} \quad {\rm
for} \quad 0 < r < \Delta \eta \; .
\end{equation}
It follows that steps 2-4 give,
\begin{eqnarray}
I_1 & = & (D\!-\!2) \times \frac{2
\pi^{\frac{D-1}2}}{\Gamma(\frac{D-1}2)} \times
\int_{-\frac1{H}}^{\eta} \!\!\! d\eta' \, (a a')^{2-\frac{D}2}
\Delta \eta \nonumber \\
& & \hspace{4cm} \times \int_0^{\Delta \eta} \!\!\! dr \, r^D
\frac{2 i \sin(\frac{\pi D}2) \, e^{-i \pi \frac{D}2} }{[\Delta
\eta^2
- r^2]^D} \; , \qquad \\
& = & (D\!-\!2) \times \frac{2
\pi^{\frac{D-1}2}}{\Gamma(\frac{D-1}2)} \times
\int_{-\frac1{H}}^{\eta} \!\!\! d\eta' \, (a a')^{2-\frac{D}2}
\Delta \eta^{-D+2} \nonumber \\
& & \hspace{4cm} \times i \sin\Bigl(\frac{D\pi}2\Bigr) e^{-i\pi
\frac{D}2}
\int_0^1 \!\! dx \, x^{\frac{D-1}2} (1 - x)^{-D} \; , \qquad \\
& = & (D\!-\!2) \times \frac{2
\pi^{\frac{D-1}2}}{\Gamma(\frac{D-1}2)}
\times H^{D-3} a \int_{\frac1{a}}^1 dt \, t^{\frac{D}2-2} (1-t)^{-D+2} \nonumber \\
& & \hspace{4cm} \times i \sin\Bigl(\frac{D\pi}2\Bigr) e^{-i\pi
\frac{D}2} \times \frac{\Gamma(\frac{D+1}2) \Gamma(-D
\!+\!1)}{\Gamma(\frac{-D+3}2)} \; , \qquad \label{I1}
\end{eqnarray}
After step four the other three integrals are,
\begin{eqnarray}
I_2 & = & \frac12 (D\!-\!4) \times \frac{2
\pi^{\frac{D-1}2}}{\Gamma( \frac{D-1}2)} \times H^{D-3} a
\int_{\frac1{a}}^1 dt \, t^{\frac{D}2-3} (1-t)^{-D+3}
\nonumber \\
& & \hspace{3cm} \times i \sin\Bigl(\frac{D\pi}2\Bigr) e^{-i\pi
(\frac{D}2-1)} \times \frac{\Gamma(\frac{D+1}2) \Gamma(-D
\!+\!2)}{\Gamma(\frac{-D+5}2)}
\; , \qquad \label{I2} \\
I_3 & = & (D\!-\!1) \times \frac{2
\pi^{\frac{D-1}2}}{\Gamma(\frac{D-1}2)}
\times H^{D-3} a \int_{\frac1{a}}^1 dt \, t^{\frac{D}2-2} (1-t)^{-D+2} \nonumber \\
& & \hspace{3cm} \times i \sin\Bigl(\frac{D\pi}2\Bigr) e^{-i\pi
(\frac{D}2-1)} \times \frac{\Gamma(\frac{D-1}2) \Gamma(-D
\!+\!2)}{\Gamma(\frac{-D+3}2)}
\; , \qquad \label{I3} \\
I_4 & = & \frac12 (D\!-\!4) \times \frac{2
\pi^{\frac{D-1}2}}{\Gamma( \frac{D-1}2)} \times H^{D-3} a
\int_{\frac1{a}}^1 dt \, t^{\frac{D}2-2} (1-t)^{-D+3}
\nonumber \\
& & \hspace{3cm} \times i \sin\Bigl(\frac{D\pi}2\Bigr) e^{-i\pi
(\frac{D}2-1)} \times \frac{\Gamma(\frac{D+1}2) \Gamma(-D
\!+\!2)}{\Gamma(\frac{-D+5}2)} \; . \qquad \label{I4}
\end{eqnarray}

\begin{table}

\vbox{\tabskip=0pt \offinterlineskip
\def\tablerule{\noalign{\hrule}}
\halign to436pt {\strut#& \vrule#\tabskip=1em plus2em& \hfil#\hfil&
\vrule#& \hfil#\hfil& \vrule#& \hfil#\hfil& \vrule#\tabskip=0pt\cr
\tablerule \omit&height4pt&\omit&&\omit&&\omit&\cr
&&$\!\!\!\!{\rm\,Term}\!\!\!\!$ && ${\rm Contribution\,\, from\,\,}
\Bigl\langle \Omega \Bigl\vert \Bigl\{
(2b),b^{\dagger}(\vec{k},s)\Bigr\} \Bigr\vert \Omega \Bigr\rangle$
&& $\!\!\!\!\!{\rm Coef.}\!\!\!\!\!$ &\cr
\omit&height4pt&\omit&&\omit&&\omit&\cr \tablerule
\omit&height4pt&\omit&&\omit&&\omit&\cr &&$\!\!\!\!{\rm
2b.II'.II'}\!\!\!\!$ && $-\frac{i \kappa^2 (D-1)}{4 (D-3)} \!
\int\!d^{D}x'
\partial_0 \partial'_0 i\delta \!\Delta^A_{\!++}\!(x;x') \gamma^0
\hspace{-.1cm}\not{\hspace{-.1cm}{\partial}} G_{\rm cf}(\!x\!-\!x')
\gamma^0 \Xi_0(x')$ &&$+0$ & \cr
\omit&height4pt&\omit&&\omit&&\omit&\cr \tablerule
\omit&height4pt&\omit&&\omit&&\omit&\cr &&$\!\!\!\!{\rm
2b.II'.II}\!\!\!\!$ && $\frac{i \kappa^2 (D-1)}{4 (D-3)} \!
\int\!d^{D}x'
\partial_0 \partial_k i\delta \!\Delta^A_{\!++}\!(x;x') \gamma^0
\hspace{-.1cm}\not{\hspace{-.1cm}{\partial}} G_{\rm cf}(\!x\!-\!x')
\gamma^k \Xi_0(x')$ &&$-\frac34$ & \cr
\omit&height4pt&\omit&&\omit&&\omit&\cr \tablerule
\omit&height4pt&\omit&&\omit&&\omit&\cr &&$\!\!\!\!{\rm
2b.II.II'}\!\!\!\!$ && $-\frac{i \kappa^2 (D-1)}{4 (D-3)} \!
\int\!d^Dx'
\partial_k \partial_0' i \delta \!\Delta^A_{\!++}\!(x;x') \gamma^k
\hspace{-.1cm}\not{\hspace{-.1cm}{\partial}} G_{\rm cf}(\!x\!-\!x')
\gamma^0 \Xi_0(x')$ &&$+\frac38$ & \cr
\omit&height4pt&\omit&&\omit&&\omit&\cr \tablerule
\omit&height4pt&\omit&&\omit&&\omit&\cr &&$\!\!\!\!{\rm
2b.II.II}\!\!\!\!$ && $\frac{i \kappa^2 (D-1)}{4 (D-3)} \!
\int\!d^Dx'
\partial_k \partial_{\ell} i\delta \!\Delta^A_{\!++}\!(x;x') \gamma^k
\hspace{-.1cm}\not{\hspace{-.1cm}{\partial}} G_{\rm cf}(\!x\!-\!x')
\gamma^{\ell} \Xi_0(x')$ &&$-\frac38$ & \cr
\omit&height4pt&\omit&&\omit&&\omit&\cr \tablerule
\omit&height4pt&\omit&&\omit&&\omit&\cr &&$\!\!\!\!{\rm
2b.II'.I}\!\!\!\!$ && $-\frac{i \kappa^2}{4 (D-3)} \! \int\!d^{D}x'
\partial_0 \partial_k i\delta \!\Delta^A_{\!++}\!(x;x') \gamma^0
\hspace{-.1cm}\not{\hspace{-.1cm}{\partial}} G_{\rm cf}(\!x\!-\!x')
\gamma^k \Xi_0(x')$ &&$+\frac14$ & \cr
\omit&height4pt&\omit&&\omit&&\omit&\cr \tablerule
\omit&height4pt&\omit&&\omit&&\omit&\cr &&$\!\!\!\!{\rm
2b.II.I}\!\!\!\!$ && $-\frac{i \kappa^2}{4 (D-3)} \! \int\!d^Dx'
\partial_k \partial_{\ell} i\delta \!\Delta^A_{\!++}\!(x;x') \gamma^k
\hspace{-.1cm}\not{\hspace{-.1cm}{\partial}} G_{\rm cf}(\!x\!-\!x')
\gamma^{\ell} \Xi_0(x')$ &&$+\frac18$ & \cr
\omit&height4pt&\omit&&\omit&&\omit&\cr \tablerule
\omit&height4pt&\omit&&\omit&&\omit&\cr &&$\!\!\!\!{\rm
2b.I.II'}\!\!\!\!$ && $\frac{i \kappa^2}{4 (D-3)} \! \int\!d^{D}x'
\partial_k \partial_0' i \delta \!\Delta^A_{\!++}\!(x;x') \gamma^k
\hspace{-.1cm}\not{\hspace{-.1cm}{\partial}} G_{\rm cf}(\!x\!-\!x')
\gamma^0 \Xi_0(x')$ &&$-\frac18$ & \cr
\omit&height4pt&\omit&&\omit&&\omit&\cr \tablerule
\omit&height4pt&\omit&&\omit&&\omit&\cr &&$\!\!\!\!{\rm
2b.I.II}\!\!\!\!$ && $-\frac{i \kappa^2}{4 (D-3)} \! \int\!d^Dx'
\partial_k \partial_{\ell} i \delta \!\Delta^A_{\!++}\!(x;x') \gamma^k
\hspace{-.1cm}\not{\hspace{-.1cm}{\partial}} G_{\rm cf}(\!x\!-\!x')
\gamma^{\ell} \Xi_0(x')$ &&$+\frac18$ & \cr
\omit&height4pt&\omit&&\omit&&\omit&\cr \tablerule
\omit&height4pt&\omit&&\omit&&\omit&\cr &&$\!\!\!\!{\rm
2b.I.I.A}\!\!\!\!$ && $-\frac{i \kappa^2}{16} \! \int\!d^Dx'
\partial_k \partial_k i\delta \!\Delta^A_{\!++}\!(x;x') \gamma^{\ell}
\hspace{-.1cm}\not{\hspace{-.1cm}{\partial}} G_{\rm cf}(\!x\!-\!x')
\gamma^{\ell} \Xi_0(x')$ &&$+\frac3{32}$ & \cr
\omit&height4pt&\omit&&\omit&&\omit&\cr \tablerule
\omit&height4pt&\omit&&\omit&&\omit&\cr &&$\!\!\!\!\!{\rm
2b.I.I.BC}\!\!\!\!\!$ && $-\frac{i \kappa^2 (D-5)}{16 (D-3)} \!
\int\!d^Dx'
\partial_k \partial_{\ell} i\delta\! \Delta^A_{\!++}\!(x;x') \gamma^k
\hspace{-.1cm}\not{\hspace{-.1cm}{\partial}} G_{\rm cf}(\!x\!-\!x')
\gamma^{\ell} \Xi_0(x')$ &&$-\frac1{32}$ & \cr
\omit&height4pt&\omit&&\omit&&\omit&\cr \tablerule
\omit&height2pt&\omit&&\omit&&\omit&\cr \tablerule
\omit&height4pt&\omit&&\omit&&\omit&\cr &&$\!\!\!\!{\rm
Total}\!\!\!\!$ && $\!\!\!\!\!-\frac{i \kappa^2}4 \!\int\!d^Dx'
\partial^{\alpha} \partial^{\prime \beta} i\delta \!\Delta^A_{\!++}\!(x;x')
[\mbox{}^{ij} T_A^{k\ell}] \gamma_i J_{j\alpha} \! \hspace{-.1cm}
\not{\hspace{-.1cm}\partial} G_{\rm cf}(\!x\!-\!x') \gamma_k J_{\ell
\beta} \Xi_{0}(x')\!\!\!\!\!$ && $\!\!\!\!\!-\frac5{16} \!\!\!\!\!$
& \cr \omit&height4pt&\omit&&\omit&&\omit&\cr \tablerule}}

\caption{The full contribution from each $(2b)$ term consists of its
numerical coefficient times $\frac{i \kappa^2 H^2}{16 \pi^2} H a
\gamma^0 \Xi_0(x)$.} \label{table2b}
\end{table}

Note that (for $D=4$) the $t$ integrands of $I_1$, $I_3$ and $I_4$
are finite at $t=0$. This means we make only an error of order $1/a$
by extending the range of $t$ down to $t=0$, at which point we get
another beta function,
\begin{eqnarray}
\int_{\frac1{a}}^1 dt \, t^{\frac{D}2-2} (1-t)^{-D+2} & = &
\frac{\Gamma( \frac{D}2 \!-\! 1) \Gamma(-D \!+\!
3)}{\Gamma(-\frac{D}2 \!+\!2)} +
O\Bigl(\frac1{a}\Bigr) \; , \\
\int_{\frac1{a}}^1 dt \, t^{\frac{D}2-2} (1-t)^{-D+3} & = &
\frac{\Gamma( \frac{D}2 \!-\! 1) \Gamma(-D \!+\!
4)}{\Gamma(-\frac{D}2 \!+\!3)} + O\Bigl(\frac1{a}\Bigr) \; .
\end{eqnarray}
This allows us to evaluate $I_1$, $I_3$ and $I_4$. Setting
$D=4-\epsilon$ and taking $\epsilon$ to zero gives the following
results for these three integrals,
\begin{eqnarray}
I_1 & \longrightarrow & 2 \times 4 \pi \times a H \times -\frac12
\times -i \frac{\pi}2 \epsilon \times \frac1{16 \epsilon} = \frac18
\pi^2 \times
i a H \; , \\
I_3 & \longrightarrow & 3 \times 4 \pi \times a H \times -\frac12
\times i \frac{\pi}2 \epsilon \times -\frac1{8 \epsilon} = \frac38
\pi^2 \times
i a H \; , \\
I_4 & \longrightarrow & -\frac12 \epsilon \times 4 \pi \times a H
\times \frac1{\epsilon} \times i \frac{\pi}2 \epsilon \times
\frac3{8 \epsilon} = -\frac38 \pi^2 \times i a H \; .
\end{eqnarray}

This procedure is {\it not} valid for the $t$ integral of $I_2$
because the integrand diverges at $t=0$. The right way to evaluate
the $t$ integral in (\ref{I2}) is to first add and subtract the $t$
integral from (\ref{I4}),
\begin{eqnarray}
\lefteqn{\int_{\frac1{a}}^1 dt \, t^{\frac{D}2-3} (1-t)^{-D+3} } \nonumber \\
& &  = \int_{\frac1{a}}^1 dt \, t^{\frac{D}2-2} (1-t)^{-D+3} +
\int_{\frac1{a}}^1 dt \, \Bigl\{\frac{1-t}{t}\Bigr\} t^{\frac{D}2-2}
(1-t)^{-D+3} \; . \qquad
\end{eqnarray}
Now extend the range in the first integral and take $D=4$ in the
second,
\begin{eqnarray}
\lefteqn{\int_{\frac1{a}}^1 dt \, t^{\frac{D}2-3} (1-t)^{-D+3} } \nonumber \\
& & = \int_0^1 dt \, t^{\frac{D}2-2} (1-t)^{-D+3} +
\int_{\frac1{a}}^1 dt \, \frac1{t} + O\Bigl(\frac1{a},D\!-\!4\Bigr)
= \frac1{\epsilon} + O(1) \; . \qquad
\end{eqnarray}
This gives,
\begin{equation}
I_2 \longrightarrow -\frac12 \epsilon \times 4 \pi \times a H \times
\frac1{\epsilon} \times i \frac{\pi}2 \epsilon \times \frac3{8
\epsilon} = -\frac38 \pi^2 \times i a H \; .
\end{equation}

\begin{table}

\vbox{\tabskip=0pt \offinterlineskip
\def\tablerule{\noalign{\hrule}}
\halign to436pt {\strut#& \vrule#\tabskip=1em plus2em& \hfil#\hfil&
\vrule#& \hfil#\hfil& \vrule#& \hfil#\hfil& \vrule#\tabskip=0pt\cr
\tablerule \omit&height4pt&\omit&&\omit&&\omit&\cr
&&$\!\!\!\!{\rm\,Term}\!\!\!\!$ && ${\rm Contribution\,\, from\,\,}
\Bigl\langle \Omega \Bigl\vert \Bigl\{
(4b),b^{\dagger}(\vec{k},s)\Bigr\} \Bigr\vert \Omega \Bigr\rangle$
&& $\!\!\!\!\!{\rm Coef.}\!\!\!\!\!$ &\cr
\omit&height4pt&\omit&&\omit&&\omit&\cr \tablerule
\omit&height4pt&\omit&&\omit&&\omit&\cr &&$\!\!\!\!{\rm
4b.II'.II'}\!\!\!\!$ && $-\frac{i \kappa^2 (D-1)}{4 (D-3)} \!
\int\!d^{D}x'
\partial_0 \partial'_0 \delta \!G_{\!\!A}\!(x;x') \gamma^0
\hspace{-.1cm}\not{\hspace{-.1cm}{\partial}} i\Delta^{\rm
cf}_{+-}(\!x\!-\!x') \gamma^0 \Xi_0(x')$ &&$+0$ & \cr
\omit&height4pt&\omit&&\omit&&\omit&\cr \tablerule
\omit&height4pt&\omit&&\omit&&\omit&\cr &&$\!\!\!\!{\rm
4b.II'.II}\!\!\!\!$ && $\frac{i \kappa^2 (D-1)}{4 (D-3)} \!
\int\!d^{D}x'
\partial_0 \partial_k \delta \!G_{\!\!A}\!(x;x') \gamma^0
\hspace{-.1cm}\not{\hspace{-.1cm}{\partial}} i\Delta^{\rm
cf}_{+-}(\!x\!-\!x') \gamma^k \Xi_0(x')$ &&$-\frac34$ & \cr
\omit&height4pt&\omit&&\omit&&\omit&\cr \tablerule
\omit&height4pt&\omit&&\omit&&\omit&\cr &&$\!\!\!\!{\rm
4b.II.II'}\!\!\!\!$ && $-\frac{i \kappa^2 (D-1)}{4 (D-3)} \!
\int\!d^Dx'
\partial_k \partial_0' \delta \!G_{\!\!A}\!(x;x') \gamma^k
\hspace{-.1cm}\not{\hspace{-.1cm}{\partial}} i\Delta^{\rm
cf}_{+-}(\!x\!-\!x') \gamma^0 \Xi_0(x')$ &&$+\frac38$ & \cr
\omit&height4pt&\omit&&\omit&&\omit&\cr \tablerule
\omit&height4pt&\omit&&\omit&&\omit&\cr &&$\!\!\!\!{\rm
4b.II.II}\!\!\!\!$ && $\frac{i \kappa^2 (D-1)}{4 (D-3)} \!
\int\!d^Dx'
\partial_k \partial_{\ell} \delta \!G_{\!\!A}\!(x;x') \gamma^k
\hspace{-.1cm}\not{\hspace{-.1cm}{\partial}} i\Delta^{\rm
cf}_{+-}(\!x\!-\!x') \gamma^{\ell} \Xi_0(x')$ &&$-\frac38$ & \cr
\omit&height4pt&\omit&&\omit&&\omit&\cr \tablerule
\omit&height4pt&\omit&&\omit&&\omit&\cr &&$\!\!\!\!{\rm
4b.II'.I}\!\!\!\!$ && $-\frac{i \kappa^2}{4 (D-3)} \! \int\!d^{D}x'
\partial_0 \partial_k \delta \!G_{\!\!A}\!(x;x') \gamma^0
\hspace{-.1cm}\not{\hspace{-.1cm}{\partial}} i\Delta^{\rm
cf}_{+-}(\!x\!-\!x') \gamma^k \Xi_0(x')$ &&$+\frac14$ & \cr
\omit&height4pt&\omit&&\omit&&\omit&\cr \tablerule
\omit&height4pt&\omit&&\omit&&\omit&\cr &&$\!\!\!\!{\rm
4b.II.I}\!\!\!\!$ && $-\frac{i \kappa^2}{4 (D-3)} \! \int\!d^Dx'
\partial_k \partial_{\ell} \delta \!G_{\!\!A}\!(x;x') \gamma^k
\hspace{-.1cm}\not{\hspace{-.1cm}{\partial}} i\Delta^{\rm
cf}_{+-}(\!x\!-\!x') \gamma^{\ell} \Xi_0(x')$ &&$+\frac18$ & \cr
\omit&height4pt&\omit&&\omit&&\omit&\cr \tablerule
\omit&height4pt&\omit&&\omit&&\omit&\cr &&$\!\!\!\!{\rm
4b.I.II'}\!\!\!\!$ && $\frac{i \kappa^2}{4 (D-3)} \! \int\!d^{D}x'
\partial_k \partial_0' \delta \!G_{\!\!A}\!(x;x') \gamma^k
\hspace{-.1cm}\not{\hspace{-.1cm}{\partial}} i\Delta^{\rm
cf}_{+-}(\!x\!-\!x') \gamma^0 \Xi_0(x')$ &&$-\frac18$ & \cr
\omit&height4pt&\omit&&\omit&&\omit&\cr \tablerule
\omit&height4pt&\omit&&\omit&&\omit&\cr &&$\!\!\!\!{\rm
4b.I.II}\!\!\!\!$ && $-\frac{i \kappa^2}{4 (D-3)} \! \int\!d^Dx'
\partial_k \partial_{\ell} \delta \!G_{\!\!A}\!(x;x') \gamma^k
\hspace{-.1cm}\not{\hspace{-.1cm}{\partial}} i\Delta^{\rm
cf}_{+-}(\!x\!-\!x') \gamma^{\ell} \Xi_0(x')$ &&$+\frac18$ & \cr
\omit&height4pt&\omit&&\omit&&\omit&\cr \tablerule
\omit&height4pt&\omit&&\omit&&\omit&\cr &&$\!\!\!\!{\rm
4b.I.I.A}\!\!\!\!$ && $-\frac{i \kappa^2}{16} \! \int\!d^Dx'
\partial_k \partial_k \delta \!G_{\!\!A}\!(x;x') \gamma^{\ell}
\hspace{-.1cm}\not{\hspace{-.1cm}{\partial}} i\Delta^{\rm
cf}_{+-}(\!x\!-\!x') \gamma^{\ell} \Xi_0(x')$ &&$+\frac3{32}$ & \cr
\omit&height4pt&\omit&&\omit&&\omit&\cr \tablerule
\omit&height4pt&\omit&&\omit&&\omit&\cr &&$\!\!\!\!\!{\rm
4b.I.I.BC}\!\!\!\!\!$ && $-\frac{i \kappa^2 (D-5)}{16 (D-3)} \!
\int\!d^Dx'
\partial_k \partial_{\ell} \delta \!G_{\!\!A}\!(x;x') \gamma^k
\hspace{-.1cm}\not{\hspace{-.1cm}{\partial}} i\Delta^{\rm
cf}_{+-}(\!x\!-\!x') \gamma^{\ell} \Xi_0(x')$ &&$-\frac1{32}$ & \cr
\omit&height4pt&\omit&&\omit&&\omit&\cr \tablerule
\omit&height2pt&\omit&&\omit&&\omit&\cr \tablerule
\omit&height4pt&\omit&&\omit&&\omit&\cr &&$\!\!\!\!{\rm
Total}\!\!\!\!$ && $\!\!\!\!\!-\frac{i \kappa^2}4 \!\int\!d^Dx'
\partial^{\alpha} \partial^{\prime \beta} \delta \!G_{\!\!A}\!(x;x')
[\mbox{}^{ij} T_A^{k\ell}] \gamma_i J_{j\alpha} \! \hspace{-.1cm}
\not{\hspace{-.1cm} \partial} i\Delta^{\rm cf}_{\scriptscriptstyle
+-}(\!x\! -\!x') \gamma_k J_{\ell \beta} \Xi_{0}(x')\!\!\!\!\!$ &&
$\!\!\!\!\! -\frac5{16}\!\!\!\!\!$ &\cr
\omit&height4pt&\omit&&\omit&&\omit&\cr \tablerule}}

\caption{The full contribution from each $(4b)$ term consists of its
numerical coefficient times $\frac{i \kappa^2 H^2}{16 \pi^2} H a
\gamma^0 \Xi_0(x)$.} \label{table4b}
\end{table}

\begin{table}
\vbox{\tabskip=0pt \offinterlineskip
\def\tablerule{\noalign{\hrule}}
\halign to200pt {\strut#& \vrule#\tabskip=1em plus2em& \hfil#\hfil&
\vrule#&  \hfil#\hfil& \vrule#\tabskip=0pt\cr \tablerule
\omit&height4pt&\omit&&\omit&\cr &&$\!\!\!\!{\rm\,Term}\!\!\!\!$ &&
$\!\!\!\!{\rm Coef.}\!\!\!\!$ &\cr \omit&height4pt&\omit&&\omit&\cr
\tablerule \omit&height4pt&\omit&&\omit&\cr
&&$\!\!\!\!{\rm\,6}\!\!\!\!$ && $0$ &\cr
\omit&height4pt&\omit&&\omit&\cr \tablerule
\omit&height4pt&\omit&&\omit&\cr &&$\!\!\!\!{\rm\,7}\!\!\!\!$ && $3$
&\cr \omit&height4pt&\omit&&\omit&\cr \tablerule
\omit&height4pt&\omit&&\omit&\cr &&$\!\!\!\!{\rm\,8}\!\!\!\!$ && $0$
&\cr \omit&height4pt&\omit&&\omit&\cr \tablerule}}

\caption{The full contribution from each term consists of its
numerical coefficient times $\frac{i \kappa^2 H^2}{16 \pi^2} H a
\gamma^0 \Xi_0(x)$.}

\label{table678}
\end{table}

Substituting our results for $I_{1-4}$ into expressions
(\ref{0,n}-\ref{mn}) gives the entries for 2b.II'.I, 2b.II.I,
2b.I.II' and 2b.I.II in Table~\ref{table2b}. Combining the totals
from Tables~\ref{table1b}-\ref{table678} gives a result in perfect
agreement with our explicit computation (\ref{keyeqn}),
\begin{eqnarray}
\lefteqn{\kappa^2 \Bigl\langle \Omega \Bigl\vert \Bigl\{ i
\hspace{-.1cm} \not{\hspace{-.10cm} \mathcal{\partial}} \Psi_2(x),
b^{\dagger}(\vec{k},s) \Bigr\} \Bigr\vert \Omega \Bigr\rangle
\longrightarrow \frac{\kappa^2 H^2}{16 \pi^2} \, i H a
\gamma^0 \Xi_0(x) } \nonumber \\
& & \hspace{1.5cm} \times \Biggl\{-\frac18 - \frac18 - \frac5{16} -
\frac5{16} + 3 \Biggr\} = \; \frac{\kappa^2 H^2}{16 \pi^2} \, i H a
\gamma^0 \Xi_0(x) \times \frac{17}8 \; . \qquad
\end{eqnarray}

\section {Discussion}
We have taken a major step in developing a technique to sum the
series of leading infrared logarithms of inflationary quantum
gravity. Our technique was to employ a previous explicit computation
\cite{MW2,MW3} as ``data'' in the search for a simple operator
formalism for reproducing the leading infrared logarithms. We found
that only gravitons with the $A$-type polarization contribute, and
only a single term (\ref{dAprop}) in the infinite series expansion
of their propagator matters. We do not yet know if our new rule
(\ref{dAsimple}-\ref{GAsimple}) reproduces the leading logarithms of
other quantities, or if it continues to work beyond one loop for the
fermion effective mode function.

One can easily see that the old rule (\ref{srule1}-\ref{srule2})
fails to reproduce the leading logarithms of quantum gravity. For
example, consider the term 1.b.II of Table~\ref{table1b},
\begin{eqnarray}
{\rm 1b.II} & \equiv & -\frac{i \kappa^2}{2 (D-3)} \!\int\!d^{D}x'
\partial_k i\delta \!\Delta^A_{\scriptscriptstyle \!++}\!(x;x')
\hspace{-.1cm}\not{\hspace{-.1cm}\overline{\partial}} \hspace{-.1cm}
\not{\hspace{-.1cm}\partial}\,G_{\rm cf}(\!x\!-\!x') \gamma^k
\Xi_{0}(x')
\; , \qquad \\
& \longrightarrow & \frac{\kappa^2 H^2}{16 \pi^2} \, i a H \gamma^0
\Xi_0(x) \times -\frac14 \; .
\end{eqnarray}
The old replacement (\ref{srule1}) corresponds to substituting a
purely temporal function for $i\delta \!\Delta^A_{\scriptscriptstyle
\!++}\!(x;x')$,
\begin{equation}
i\delta \!\Delta^A_{\scriptscriptstyle \!++}\!(x;x') \longrightarrow
\frac{H^{D-2}}{(4 \pi)^{\frac{D}2}}
\frac{\Gamma(D\!-\!1)}{\Gamma(\frac{D}2)} \, 2 \ln\Bigl[{\rm
min}(a,a') \Bigr] \; .
\end{equation}
Of course the spatial derivative of this would give {\it zero} for
1b.II!

There are also problems with the closely related contribution from
3b.II,
\begin{eqnarray}
{\rm 3b.II} & \equiv &-\frac{i \kappa^2}{2 (D-3)} \int\!d^Dx'
\partial_k
\delta \!G_{\!\!A}\!(x;x') \hspace{-.1cm} \not{\hspace{-.1cm}
\overline{\partial}} \hspace{-.1cm}\not{\hspace{-.1cm}{\partial}} \,
i\Delta^{\rm cf}_{\scriptscriptstyle \!+-}\!(x\!-\!x') \gamma^k
\Xi_{0}(x')
\; , \qquad \\
& \longrightarrow & \frac{\kappa^2 H^2}{16 \pi^2} \, i a H \gamma^0
\Xi_0(x) \times -\frac14 \; .
\end{eqnarray}
The old replacement (\ref{srule2}) corresponds to the substitution,
\begin{equation}
\delta \!G_{\!\! A}\!(x;x') \longrightarrow \frac{\theta(t\!-\!t')
\delta^{D-1}(\vec{x} \!-\! \vec{x}')}{(D\!-\!1) H a^{\prime D-1}} \;
.
\end{equation}
This gives a nonzero result, but not the right one,
\begin{eqnarray}
\lefteqn{-\frac{i \kappa^2}{2 (D-3)} \int\!d^Dx' \partial_k \Biggl\{
\frac{\theta(t\!-\!t') \delta^{D-1}(\vec{x} \!-\!
\vec{x}')}{(D\!-\!1) H a^{\prime D-1}} \Biggr\} \hspace{-.1cm}
\not{\hspace{-.1cm} \overline{\partial}}
\hspace{-.1cm}\not{\hspace{-.1cm}{\partial}} \, i\Delta^{\rm
cf}_{\scriptscriptstyle \!+-}\!(x\!-\!x') \gamma^k \Xi_{0}(x') }
\nonumber \\
& & \hspace{3cm} \longrightarrow \frac{i \kappa^2}{2 \pi^{\frac{D}2}
H} \, \frac{\Gamma(\frac{D}2 \!+\! 1)}{D \!-\! 3} \, \gamma^0
\Xi_0(x) \int_{-\frac1{H}}^{\eta}
\frac{d\eta'\,e^{i\pi\frac{D}{2}}}{a^{\prime D-1} \Delta \eta^{D+1}}
\; , \qquad \\
& & \hspace{3cm} \longrightarrow \frac{\kappa^2 H^2}{16 \pi^2} \, i
a H \gamma^0 \Xi_0(x) \times -4 \; .
\end{eqnarray}

In fact the old rule (\ref{srule1}-\ref{srule2}) does not give
correct results for any of the {\it thirty} distinct nonlocal
contributions of Tables~\ref{table1b}-\ref{table4b}! The failure of
this rule --- which works for models without derivative couplings
\cite{MW1,SY,PTsW3}
---deserves comment. Massless, minimally coupled scalars and gravitons
are active fields. In order to produce infrared logarithms a theory
must possess interactions involving at least one undifferentiated
active field. However, there is a hierarchy of increasingly
complicated ways in which this can happen:
\begin{enumerate}
\item{The theory may involve {\it only} undifferentiated active fields;}
\item{The theory may involve active and passive fields with non-derivative
interactions; and}
\item{The theory may involve differentiated active fields, with or without
passive fields.}
\end{enumerate}
The relation of our new rule (\ref{dAsimple}-\ref{GAsimple}) to the
old rule (\ref{srule1}-\ref{srule2}) can be understood by
considering how the expectation value of a given term in the free
field expansion of some operator attains leading logarithm order in
each case.

When only undifferentiated active fields are present, each pair of
free fields and each vertex integration must contribute to an
infrared logarithm \cite{PTsW3}. Therefore only the infrared part of
the free field mode sum matters and one can effect this truncation
at the level of the Yang-Feldman equations. This is the case solved
by Starobinski\u{\i} and Yokoyama \cite{SY}. At the level of
expectation values of the free field expansion it corresponds to the
replacements (\ref{srule1}-\ref{srule2}) with $D=4$ because there
are no ultraviolet divergences at leading logarithm order.

When passive fields are present, but the active fields are not
differentiated, reaching leading logarithm order still requires
every active field or active Green's function to contribute to an
infrared logarithm. Passive fields cannot be infrared truncated
because the order $[\ln(a)]^0$ contributions they make derive from
all parts of the free field mode sum. However, {\it precisely
because passive fields cannot produce infrared logarithms} we can
perform passive vertex integrations without accounting for the
spacetime dependence of active fields or Green's functions. This
amounts to integrating out the passive fields and then evaluating
the resulting, nonlocal effective action assuming the active fields
are constant--- which defines the effective potential. At the level
of expectation values of the free field expansion it corresponds to
the replacements (\ref{srule1}-\ref{srule2}) with dimensional
regularization on because there can be ultraviolet divergences at
leading logarithm order.

The situation is vastly more complicated when differentiated active
fields are present. In this case the {\it vertex integration} of a
differentiated active field propagator or Green's function can
produce an infrared logarithm, even though the integrand contains no
logarithm. For the fermion wave function, {\it every} infrared
logarithm arises in this fashion. We cannot ignore differentiated
active fields because they can still contribute infrared logarithms.
Neither can we ignore their spacetime dependence in performing
vertex integrations, and we must retain dimensional regularization
in order to define these integrals. In view of this it seems
doubtful that any infrared truncated formalism can correctly
represent the theory, even at leading logarithm order. The
replacements (\ref{dAsimple}-\ref{GAsimple}) of our new rule seem to
represent the appropriate generalizations of the old rule
(\ref{srule1}-\ref{srule2}) to this more singular environment.

In addition to showing that the new rule works, our analysis
provides a deeper understanding of why the fermion mode function
acquires a secular enhancement whereas the scalar mode function does
not \cite{KW2}. The reason is spin. At late times the kinetic
energies of all quanta redshift to zero. This is why we could
neglect the $\partial'_{\rho} \gamma_{\sigma} \Xi_0(x')$
contributions from terms $(1a)$, $(2a)$, $(3a)$ and $(4a)$ of
Table~\ref{contribs}. A massless scalar interacts with gravity only
through its kinetic energy. Inflationary particle production
immerses such a scalar in a sea of infrared gravitons but they do
little because the interaction is so weak. In contrast, a massless
fermion possesses an additional gravitational interaction through
its spin, which does not redshift. That is why we found leading
order contributions from the $\gamma_{\rho} J_{\sigma \beta}
\Xi_0(x')$ terms of $(1b)$, $(2b)$, $(3b)$ and $(4b)$ on
Table~\ref{contribs}.

Gravitons also have spin and it is natural to wonder what the sea of
infrared gravtions does to itself. One could answer this by using
the known one loop graviton self-energy \cite{TW0} to correct the
graviton mode functions, just as we have done for fermions. It would
also be interesting to understand in this way the null result that
has been obtained at one loop order for the graviton 1-point
function \cite{TW6}. In particular, can the spin-spin interaction
lead to significant quantum gravitational back-reaction?

\begin{center}
{\bf Acknowledgements}
\end{center}

The authors are grateful to E. O. Kahya, T. Prokopec and N. C.
Tsamis for discussions on the fascinating problem of summing the
leading infrared logarithms of quantum gravity. This work was
partially supported by FOM grant FOM-07.0583, by the Institute for
Theoretical Physics of Utrecht University, by NSF grant PHY-0653085,
and by the Institute for Fundamental Theory at the University of
Florida.

\end{document}